\documentclass[%
 reprint,
superscriptaddress,
 amsmath,amssymb,
 aps,
prb,
]{revtex4-2}
\newcommand{\bra}[1]{\langle#1\rvert} 
\newcommand{\ket}[1]{\lvert#1\rangle} 
\newcommand{\qprod}[2]{ \langle #1 | #2 \rangle} 

\usepackage{graphicx}
\usepackage{dcolumn}
\usepackage{bm}
\usepackage[hidelinks]{hyperref}
\usepackage[dvipsnames]{xcolor}
\usepackage{colortbl}
\hypersetup{
    colorlinks,
    linkcolor={red!50!black},
    citecolor={blue!50!black},
    urlcolor={blue!80!black}
}

\begin{document}

\preprint{sDFT}

\title{Stochastic and Mixed Density Functional Theory within the projector augmented wave formalism for the simulation of warm dense matter}

\author{Vidushi Sharma}
\affiliation{Theoretical Division, Los Alamos National Laboratory, Los Alamos, NM 87545, USA}
\affiliation{Center for Nonlinear Studies, Los Alamos National Laboratory, Los Alamos, NM 87545, USA}
\author{Lee A. Collins}
\author{Alexander J. White}
\email{alwhite@lanl.gov}
\affiliation{Theoretical Division, Los Alamos National Laboratory, Los Alamos, NM 87545, USA}

\noaffiliation

\date{\today}

\begin{abstract}
    Stochastic and mixed stochastic-deterministic density functional theory (DFT) are promising new approaches for the calculation of the equation-of-state and transport properties in materials under extreme conditions. In the intermediate warm dense matter regime, a state between correlated condensed matter and kinetic plasma, electrons can range from being highly localized around nuclei to delocalized over the whole simulation cell. The plane-wave basis pseudo-potential approach is thus the typical tool of choice for modeling such systems at the DFT level. Unfortunately, the stochastic DFT methods scale as the square of the maximum plane-wave energy in this basis. To reduce the effect of this scaling, and improve the overall description of the electrons within the pseudo-potential approximation, we present stochastic and mixed DFT developed and implemented within the projector augmented wave formalism. We compare results between the different DFT approaches for both single-point and molecular dynamics trajectories and present calculations of self-diffusion coefficients of solid density carbon from 1 to 50 eV.   
    
\end{abstract}

\maketitle

The warm, dense matter  (WDM) regime encompasses a wide variety of extreme environments and provides an excellent testing ground for methods that determine the basic properties of matter. These environments include, for example, planetary interiors \cite{nas-ess-2018,bethkenhagen-848-2017, Teanby2020}, stellar systems such as brown and white dwarfs \cite{becker-156-2018}, and the capsule compression stage in inertial confinement fusion (ICF) \cite{hu-25-2018, fenandez-122-2019, Bachmann2022}.  In the ice giant planets, the interiors may support a superionic phase in which hydrogen remains fluid within the lattices of the heavier constituents such as oxygen, carbon, nitrogen, and silicon \cite{millot-14-2018,cheng-17-2021,gao-128-2022}, which may help explain the anomalous planetary magnetic fields of Neptune and Uranus. In addition, the difference between an exothermic Neptune and an endothermic Uranus may originate in the nucleation of diamonds from hydrocarbon mixtures \cite{cheng-arXiv-2023,ross-292-1981}. Finally, properties of various hydrocarbons such as equations-of-state and thermal conductivities determine the performance of ICF capsules irradiated by laser pulses \cite{hu-25-2018}, and stopping power characterizes the cooling effects of deposition of the capsule material into the hydrogen fuel \cite{white-98-2018,ding-121-2018}. The activation of the James Webb Space Telescope (JWST)  presages an explosion in discoveries \cite{ahrer-arX-2022}  of exoplanets representing a vast range of physical conditions, distributions, and dynamics involving, to list just a few, surface-atmosphere couplings \cite{detrich-517-2022}, interfaces between solid and liquid components of interiors \cite{miyazaki-3-2022}, and formation pathways \cite{liu-542-2019}. In another area, the breakthrough fusion milestone \cite{zylstra-601-2022} at the National Ignition Facility emphasizes the role played in modeling by ever-improved basic physical attributes.  Both of these developments signal a pressing need for more accurate static and dynamical microscopic properties over a broad range of WDM conditions.

Many methods exist to determine the basic structure and dynamics of WDM; the most accurate arise from first-principles (FP) techniques such as density functional theory (DFT) \cite{kohn-140-1965, burke2014, Bonitz2020} and path integral Monte Carlo \cite{ceperley-67-1995, militzer-108-2012}, which supply a consistent set of basic material properties as equation-of-state, opacities, mass transport, electrical and thermal conduction. Recently, results from DFT simulations have provided training information to determine model potentials from machine learning techniques (MLP) \cite{deringer-31-2019,cheng-arXiv-2023,gao-128-2022,li-128-2022}. Kohn-Sham (KS) DFT combined with the plane-wave pseudo-potential (PWPP) method is the theory of choice for studying the electronic structure of numerous materials, ranging from solid-state condensed matter to hot dense plasmas. The success of DFT stems from the balance between computational complexity and useful accuracy, achieved by replacing the quantum-mechanical wavefunction by a much simpler quantity: the KS density matrix, typically constructed from the KS Hamiltonian eigenstates.

The cubic scaling of the computational complexity of KS-DFT with respect to system size and temperature is a major limitation \cite{Blanchet_2020}. For WDM systems, orbital-free DFT has been a particularly useful alternative, but it is based on an approximate treatment of the electron non-interacting kinetic energy \cite{Lambert_2006, Ticknor_2016, White_2017}. Linear scaling methods, such as stochastic DFT (sDFT) \cite{Fabian2019}, provide a full KS accuracy alternative for large or hot systems \cite{Cytter2018}. The more general mixed stochastic-deterministic DFT (mDFT) shows great promise for providing full KS-DFT accuracy for calculations at any temperature \cite{WhiteCollins2020}. 
However, when combined with PWPP method, sDFT, and by extension mDFT, has a quadratic dependence of the computational cost on the maximum plane-wave energy (E$_{cut}$), \emph{i.e.}, on the grid resolutions, compared to standard deterministic DFT's linear dependence. Moreover it has only been formulated in combination with norm-conserving pseudopotentials, which typically show either low accuracy or require higher E$_{cut}$. Maintaining high accuracy and low E$_{cut}$, soft pseudopotentials, requires utilization of a non-orthogonal basis, as first developed by Vanderbilt \cite{Laasonen_1993}. 

The projector augmented wave (PAW) approach, first introduced by Bl{\"o}chl \cite{Blochl_1994} and then reformulated by Kresse \cite{Kresse_1999}, generalizes the soft pseudopotentials to a formally ``all-electron" formalism. The PAW method provides a realistic description of core electrons, has a long and continued history of development, and yields accuracy comparable to more expensive ``all-electron" methods \cite{Lejaeghere_2016}. Moreover, it allows for very low E$_{cut}$, suitable for sDFT calculations. In this letter, we develop mDFT, and sDFT by limitation, within the PAW formalism and present isochoric calculations and analysis for warm dense carbon spanning the WDM regime, 1 to 50 eV.   

While DFT was initially developed for electrons in their ground-state, \emph{i.e.}, zero temperature ($\text{T}\to 0$), the temperatures in WDM systems are of the order of the Fermi energy and thus require the application of a finite-temperature formulation of KS-DFT. Mermin's formulation of DFT within the grand canonical ensemble is the most common approach used in WDM \cite{Mermin1965}. In this formulation, the single-particle KS eigenstates are partially occupied according to the Fermi-Dirac distribution function (here assuming paired spin):
\begin{align}
    f(\varepsilon) &= \frac{2}{1+e^{(\varepsilon-\mu)/k_B T}} ~,
\end{align} 
where $\mu$ is the chemical potential, and $k_B$ is the Boltzmann constant. Thus the thermal density matrix,  $\widehat \rho = f(\widehat{H}_{\text{KS}})$, is constructed as:
\begin{align}
    \widehat \rho = \sum_{b} f(\varepsilon_b) \ket{\psi_b}\bra{\psi_b}~,
\end{align}
where $\psi_b$ is an eigenvector of the KS Hamiltonian, $\widehat{H}_{\text{KS}}$, with eigenenergy $\varepsilon_b$.
In finite-temperature metals and plasmas (where electrons populate the conduction band) the number of states required to resolve all the electrons grows as $VT^{3/2}$, where $V$ is the system size, leading to cubic computational scaling in both size and temperature. 

To address this drawback of traditional KS-DFT, Baer \textit{et al.} proposed an alternative algorithmic approach to DFT that is stochastic in nature (sDFT) \cite{Baer2013, Neuhauser2014, Cytter2018, Fabian2019, Baer2022}.
In contrast to traditional KS-DFT, sDFT scales as $V/T$ and the operations on the stochastic vectors are trivial to parallelize.
sDFT is based on  Hutchinson's stochastic trace estimation (STE) \cite{Hutchinson1990} and the stochastic projection for matrices, \emph{i.e.}, it is the application of STE to the Kohn-Sham density matrix. In sDFT, the thermal density matrix, $\widehat \rho = f(\widehat{H}_{\text{KS}})$, is projected onto $N_{\chi}$ stochastic vectors ($\chi_a$):
\begin{align}
    \widehat \rho = \sum_{a\in N_\chi} f^{\frac{1}{2}}(\widehat{H}_{\text{KS}}) \ket{\chi_a}\bra{ \chi_a} f^{\frac{1}{2}}(\widehat{H}_{\text{KS}})~,
\end{align}
rather than on KS eigenstates. A converged calculation, with respect to $N_\chi$ has the same exact accuracy as that of a traditional KS-DFT calculation based on finding eigenstates.



\begin{figure}[t]
    \centering
    \includegraphics[width=3.5in]{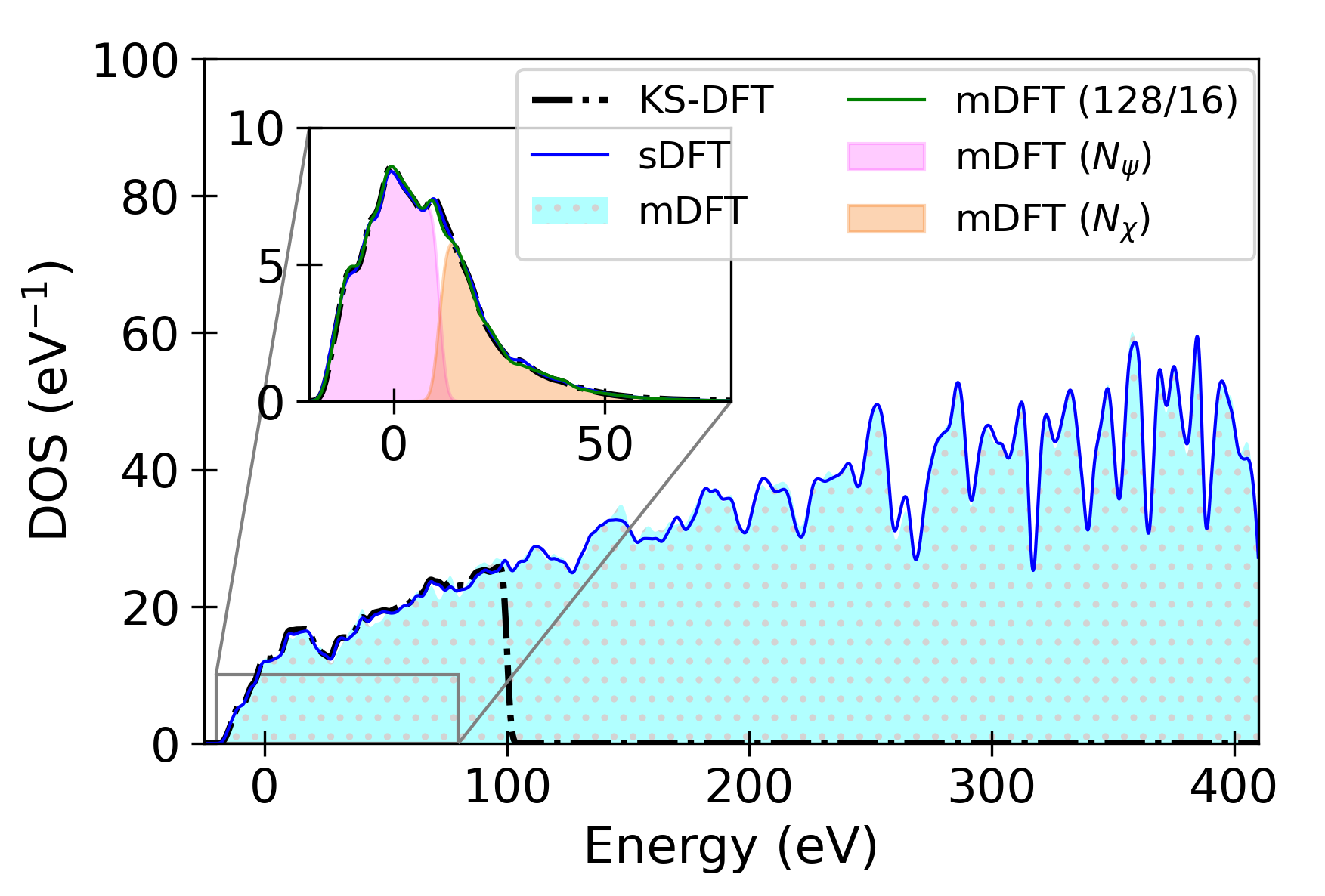}
    \caption{Disordered 64 carbon atoms system at $(\rho, \text{T}) = (3.52\,\, \text{g/cc}, 10\,\, \text{eV})$: Density of states (DOS), and (inset) Occupied DOS, obtained with Kohn-Sham (KS-DFT), stochastic (sDFT), and mixed (mDFT) methods. The chemical potential of the system is $\mu = 7.92$ eV. The pink and orange--shaded regions in the inset indicate the splitting due to deterministic ($N_\psi$) and stochastic ($N_\chi$) treatments in mDFT.}
    \label{fig:fig1}
\end{figure}
Recently, White and Collins \cite{WhiteCollins2020} proposed the mDFT approach that generalizes stochastic and deterministic KS-DFT approaches and improves the computational complexity over a wide range of temperatures. It is based on partitioning the full eigenspectrum of $\widehat{H}_{\text{DFT}}$ into low-energy and high-energy segments such that the maximally occupied low-energy eigenstates ($\psi$) are explicitly resolved while the higher-energy states are spanned by random stochastic vectors ($\chi'$).
 That is:
\begin{align}
    \ket{\chi_a'} =  \Big(\widehat I - \sum_{b\in N_\psi} \ket{\psi_b}\bra{\psi_b}\Big)\ket{\chi_a}  
\end{align}
where $\chi=e^{i 2 \pi \vec{\theta}}/N$ and $\theta \in \{0,1\}$ is a set of uncorrelated random numbers for each basis function and $N$ is a normalization constant. We define ``occupied'' stochastic vectors, ${X'} = f^{\frac{1}{2}}(\widehat{H}_{\text{DFT}}){\chi'}$, to obtain the mixed density matrix as,
\begin{align}
    {\widehat\rho } = \sum_{a\in N_\chi} \ket{{X}_a'}\bra{{X}_a'} + \sum_{b\in N_\psi} \ket{\psi_b} f(\varepsilon_b) \bra{\psi_b} ~.
\end{align}
 All observables can be expressed as traces over appropriate operators with this density matrix. See \cite{WhiteCollins2020} for a detailed description.

Figure \ref{fig:fig1} shows the density of states (DOS) and occupied DOS for a disordered carbon system, obtained with KS-DFT ($N_\psi=1024$), sDFT ($N_\chi=256$) and mDFT ($N_\psi/N_\chi=128/16$) methods. The low-energy deterministic component of mDFT is shown in pink and there is an overall good agreement across the three methods. The overlap of the components is due to the finite width of Gaussian functions used to define the continuous DOS.



The state-of-the-art for balancing accuracy, computational complexity, and grid resolution/plane-wave count is the pseudo-augmented wave (PAW) method. In principle, PAW is an all-electron method, assuming a complete set of partial waves and projectors. However, only a finite set is used in practice. 
The PAW method is based on a linear transformation matrix ($\widehat{\tau}$) connecting the smooth pseudo density matrix ($\widetilde{\rho}\,$) to an all-electron density matrix ($\widehat \rho \,$),
\begin{align}
    \widehat{\rho} &= \widehat{\tau}{\,\widetilde{\rho}\,\,}{\widehat\tau}^\dagger  
    \\
    \widehat{\tau} &= \widehat{I} + \sum_{i} \big(\ket{\phi_i} - \ket{\tilde{\phi}_i}\big) \bra{p_i}~,
\end{align}
with $\qprod{p_i}{\tilde{\phi_j}}=\delta_{ij}$. Here, $\ket{\phi_i}$ is a `true' all-electron partial wave and $\ket{\tilde{\phi_i}}$ is a pseudo partial wave dual to the projector $\ket{p_i}$. This transformation is exact in the limit of a complete set of partial waves/projectors. These functions are defined in an ``augmentation sphere" around an atom. Expectation values are preserved by defining pseudized operators, $\widetilde O$ as:
\begin{align}
    E[\widehat{O}] = Tr[\widehat{\rho} \, \widehat{O} ] = Tr[\widetilde{\rho} \, \widetilde{O} ] \,, \, \text{with} \,  \widetilde{O}=\widehat \tau^\dagger {\widehat O} \widehat \tau
\end{align}

The transformed identity operator gives an S-orthogonality condition for the transformed wavefunctions:
\begin{align}
    \label{Sop}
    &\widehat{S} = \widehat{\tau}^\dagger\widehat{\tau} =   \widehat{I} + \sum_{i,j} \ket{p_i} \big(\qprod{\phi_i}{\phi_j} - \qprod{\tilde{\phi}_i}{\tilde{\phi}_j}\big) \bra{p_j} ~,\\
    &\qprod{\psi_a}{\psi_b} = \bra{\tilde{\psi}_a}\widehat{S}\ket{\tilde{\psi}_b} = \delta_{ab} ~.
\end{align}
Therefore, $\tilde \psi_b$ is a solution to the generalized eigenvalue problem,  $\widetilde H_{\text{KS}} \tilde \psi = \varepsilon \widehat{S} \tilde \psi$. This S-orthogonality condition complicates the generation of transformed stochastic vectors. 

Our approximate projection via all-electron norm-conserving and transformed stochastic vectors is given by: 
\begin{equation}
\label{stochI}
    \widehat{I} \approx \sum_{a} \ket{\chi_a}\bra{\chi_a} = \sum_{a} \widehat{\tau}\ket{\tilde \chi_a}\bra{\tilde \chi_a}\widehat{\tau}^\dagger ~.
\end{equation}
From the transformation of the stochastic vectors and the identity operator, Eq. \eqref{stochI}, and the $\widehat S$ operator, Eq. \eqref{Sop}, we find that
\begin{align}\nonumber
    \sum_{a} \widehat{\tau}^\dagger\widehat{\tau}\ket{\tilde{\chi}_a}\bra{\tilde{\chi}_a}\widehat{\tau}^\dagger\widehat{\tau} &\approx \widehat{S} ~,\\
    \sum_{a} \ket{\tilde{\chi}_a}\bra{\tilde{\chi}_a} \approx  \sum_{b} \ket{\tilde \psi_b} \bra{\tilde \psi_b} &= \widehat{S}^{-1}  ~.
\end{align}
We now define a set of stochastic vectors $\ket{\bar{\chi}_a}$ such that
\begin{equation*}
    \widehat{I} \approx \sum_{a}\ket{\bar{\chi}_a}\bra{\bar{\chi}_a} ~, ~ \text{i.e.} ~ \delta({\vec r},{\vec r\,}') = \sum_{a} e^{i 2 \pi \left({\vec \theta}_a({\vec r})-{\vec \theta}_a({\vec r\,}')\right)}/N^2 ~,
\end{equation*}
which has the same form as the all-electron stochastic vectors, but with ${\vec r}$ and ${\vec r \,'}$ from the coarser grid of the transformed functions \cite{Cytter2018}.
Using the identity $\widehat{I} = \widehat{\tau}\widehat{S}^{-1}\widehat{\tau}^\dagger$ (see Supplementary Information \cite{SIcite}, Eq. S8) and Eq. \eqref{stochI}, we obtain
\begin{align}
    \nonumber
    \widehat{I} \approx \sum_{a} \widehat{\tau}\widehat{S}^{-\frac{1}{2}}\ket{\bar{\chi}_a}\bra{\bar{\chi}_a}\widehat{S}^{-\frac{1}{2}}\widehat{\tau}^\dagger &= \sum_{a} \widehat{\tau}\ket{\tilde{\chi}_a}\bra{\tilde{\chi}_a}\widehat{\tau}^\dagger  ~,\\
    \text{where } \quad \ket{\tilde{\chi}_a} &= \widehat{S}^{-\frac{1}{2}}\ket{\bar{\chi}_a} ~.
\end{align}
An efficient and sufficiently accurate procedure for applying $\widehat{S}^{-\frac{1}{2}}$ to vectors was formulated recently by Li \& Neuhauser \cite{Neuhauser2020}. Using the same identity, the pseudized density matrix can be written as (see Supplementary Information \cite{SIcite}):
\begin{align}
    \widetilde \rho =  f(\widehat{S}^{-1} \widetilde{H}_{\text{KS}}) \widehat{S}^{-1} = \\\nonumber f^{\frac{1}{2}}(\widehat{S}^{-1} \widetilde{H}_{\text{KS}}) \widehat{S}^{-1} f^{\frac{1}{2}}( \widetilde{H}_{\text{KS}} \widehat{S}^{-1}) ~.
\end{align}

The procedure for PAW mDFT is similar to the all-electron or norm-conserving case \cite{WhiteCollins2020} with three modifications: (i) the orthogonal stochastic vectors, $\bar \chi$, are generated and rotated to the standard PAW frame, $\tilde \chi$, (ii) the generalized eigenvalue problem is iteratively solved to obtain $\tilde \psi$, (iii) the projection of the eigenstates from the stochastic vectors is performed via: 
\begin{align}
    \ket{\tilde \chi_a'} =  \Big(\widehat S^{-1} - \sum_{b\in N_\psi} \ket{\psi_b} \bra{\psi_b}\Big) \widehat{S} \ket{\tilde \chi_a}&, \, \text{giving} \\
    \widehat S^{-1} \approx \sum_{a \in N_\chi} \ket{\tilde{\chi}_a'}\bra{\tilde{\chi}_a'} +  \sum_{b \in N_\psi} \ket{\tilde \psi_b} \bra{\tilde \psi_b}& \, \, \text{and} \\\label{DMpaw} 
    {\widetilde \rho } = \sum_{a\in N_\chi} \ket{{\tilde X}_a'}\bra{{\tilde X}_a'} + \sum_{b\in N_\psi} \ket{\tilde \psi_b} f(\varepsilon_b) \bra{\tilde \psi_b}&\, , \, \text{with} \\\label{filterpaw} ~
    \ket{{\tilde X}_a'} = f^{\frac{1}{2}}(\widehat{S}^{-1} \widetilde{H}_{\text{KS}}) \ket{{\tilde \chi}_a'}&.
\end{align}
In Eq. \eqref{filterpaw}, the $S^{-1}$ can be applied via the Woodbury formula \cite{LEVITT201598}. From Eq. \eqref{DMpaw}, all observables necessary to complete the PAW formalism, \emph{e.g.}, the on-site density matrix and the compensation charge density, can be calculated \cite{Torrent_2008}. The generalization of PAW force and stress tensor contributions for mDFT/sDFT are presented in Supplementary Information S2 \cite{SIcite}.

\begin{figure}[t]
    \centering
    \includegraphics[width=3.5in]{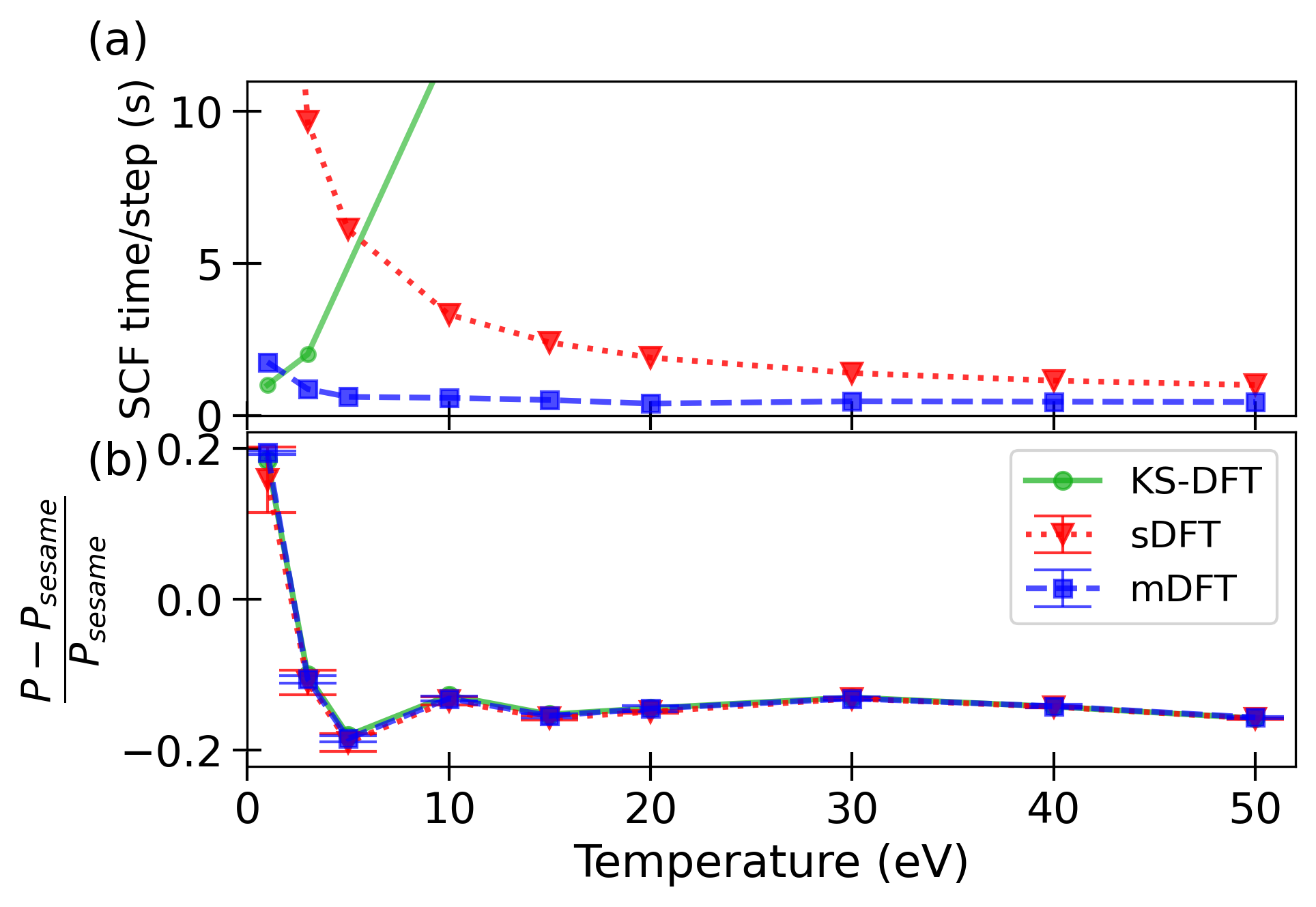}
    \caption{Disordered carbon system comprising 64 atoms at $\rho=3.52$ g/cc. Comparison of (a) SCF times per cycle, and (b) relative pressure with reference to \texttt{SESAME 7833} ($P_{\text{sesame}}$) \cite{sesame} obtained for deterministic (Kohn-Sham), stochastic, and mixed DFT calculations performed using a Cray compilation of \texttt{SHRED} on 128 cores.}
    \label{fig:fig2}
\end{figure}

To test our method, we first perform single-point ground-state energy calculations on a 64-atom disordered carbon system at several temperatures and a solid-state density of 3.52 g/cc, using the $4e^-$ PAW potential.
All the computations are using our plane-wave DFT code, \texttt{SHRED} (Stochastic and Hybrid Representation Electronic Structure by Density Functional Theory) that relies on a modified version of a portable PAW library \texttt{LibPAW} \cite{libpaw2016}, developed under the \texttt{ABINIT} \cite{abinit2020} project, and \texttt{LibXC} \cite{libxc2018} for the exchange-correlation energy functionals.
A kinetic energy cutoff E$_{cut}$ of 426 eV $(N_{\text{grid}} = 48^3)$ is used for the transformed functions and 758 eV $(N_{\text{grid}} = 64^3)$ for the densities and the spherical PAW grid around each atom.

Figure \ref{fig:fig2} shows a comparison of the three DFT algorithms in terms of computational time per self consistent field (SCF) cycle on 128 CPUs, and the accuracy and precision of pressure (for this single-point calculation) referenced to \texttt{SESAME 7833} value \cite{sesame}. The free energy, pressure and chemical potential for each temperature, along with 
the combination of orbitals $N_\psi/N_\chi$ ($N_\psi$) used in mDFT are listed in Supplementary Information, Table S1 \cite{SIcite}. The combinations range from $136/4$ at $k_B T=1$ eV to $64/40$ at $k_B T=50$ eV; whereas sDFT times are computed with $N_\chi=128$ orbitals for all temperatures. The pressures in sDFT and mDFT, are computed as averages over 10 independent SCF runs using a different set of $N_{\chi}$ stochastic orbitals; with the statistical error (the error bars) expressed as the standard deviation of the sample.
The nonlinear nature of the SCF cycle leads to a potential bias in sDFT/mDFT given by the difference between the expected value and the KS-DFT result. Mixed DFT yields energies to within 0.2\% (standard deviation of 0.3\%) of the reference KS-DFT values with a $42\times$ speedup as compared to KS-DFT at $\text{T}=20$ eV.
Additionally, the chemical potential and pressure are converged to 0.13 eV (standard deviation of 0.18 eV) and 7.27 GPa (standard deviation of 8.29 GPa) relative to their respective reference KS-DFT results. 
 

\begin{figure}[t]
    \centering
    \includegraphics[width=3.2in]{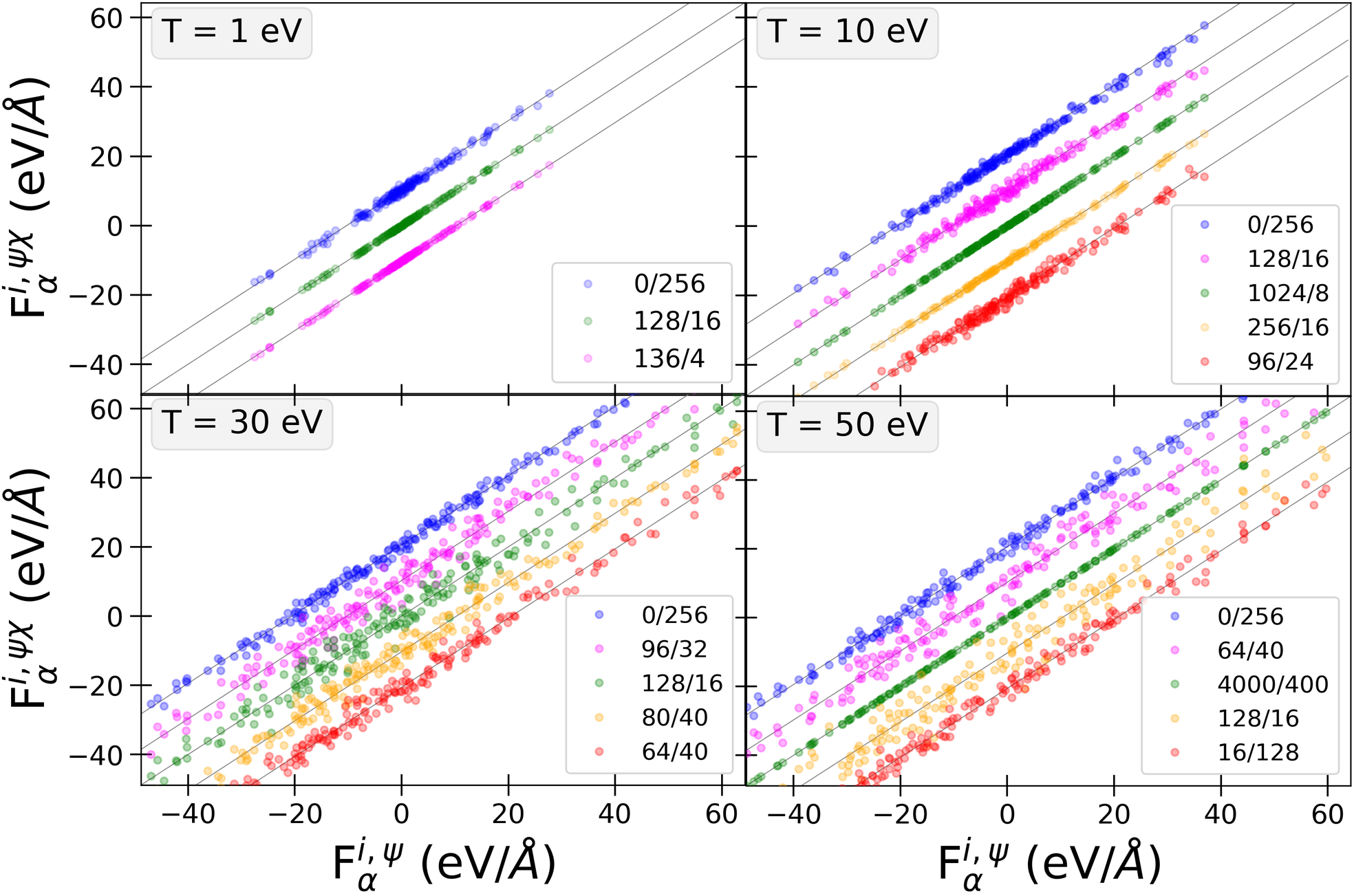}
    \caption{Comparison of mixed (F$^{i, \psi\chi}_{\alpha}$) vs Kohn-Sham (F$^{i, \psi}_{\alpha}$) DFT components of forces on all atoms obtained for various $N_\psi/N_\chi$. The agreement between stochastic (0/256) and deterministic forces improves at higher temperatures. The data points shown in magenta represent the chosen $N_\psi/N_\chi$ for mixed DFT calculations at a given temperature (T). At higher temperatures, the area of the force plots is zoomed in to keep a constant scale. The order of lines at each T matches the key.}
    \label{fig:fig3}
\end{figure}

\begin{figure}[t]
    \centering
    \includegraphics[width=2.7in]{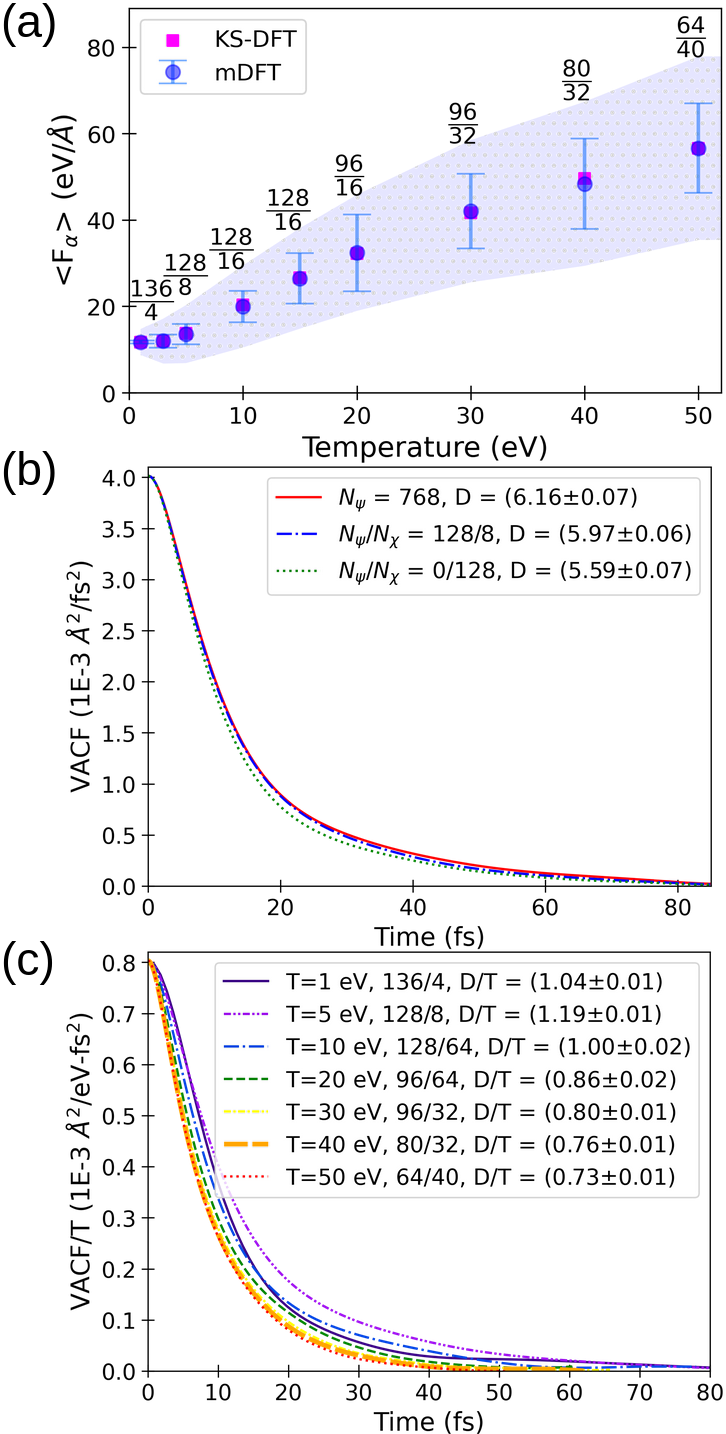}
    \caption{Disordered 64 carbon atoms system at $\rho=3.52$ g/cc: average magnitude of force on atoms $\langle \text{F}_{\alpha}\rangle$ obtained with Kohn-Sham and mixed DFT for a single snapshot. For KS-DFT, $N_\psi$ ranges from 256 at T = 1 eV to 6400 at T = 50 eV. The error bars indicate the statistical error over mixed DFT runs, and the shaded region represents a Langevin-type friction term at the respective temperature, $\langle \text{F}_{\alpha}\rangle\pm2\varsigma$ with $\gamma_\alpha=0.04$ fs$^{-1}$ \cite{Arnon2020}. A comparison of (b) velocity autocorrelation function (VACF) for KS (solid line), mixed (dash-dot line) and stochastic (dotted line) DFT methods at T=5 eV, and (c) VACF/T at different temperatures (T). The self-diffusion coefficients (D, integral of VACF) are given in the key with units of $10^{-3}$ cm$^2$/s.}
    \label{fig:fig4}
\end{figure}
Local quantities such as the electronic forces on nuclei depend on the electronic density and hence do not benefit from the self-averaging effect of stochastic DFT \cite{Baer2013}. We examine the stochastic and mixed DFT forces for several $N_\psi/N_\chi$ at different temperatures. A comparison is presented in Fig. \ref{fig:fig3}, where $F ^{i, \psi\chi}_{\alpha, n}$ indicates mixed or stochastic forces and F$^{i, \psi}_{\alpha}$ indicates deterministic KS- forces, such that $i= \{x,y,z\}$, $\alpha=1,\dots,N_{at}$ indexes the atoms, and $n=1,\dots,10$ indexes the stochastic/mixed run. The absence of an $i$ index indicates the magnitude of the force, lack of $n$ indicates the value is averaged over the 10 independent stochastic/mixed runs, and the absence of $\alpha$ indicates average over atoms. The standard deviation over $n$ independent runs, averaged over the atoms, is given by: 

\begin{align}
\sigma^{\psi\chi} &= \sum_{\alpha=1}^{N_{at}} N_{at}^{-1}   \sqrt{ \sum_{n=1}^{10} \frac{( \text{F} ^{\psi\chi}_{\alpha, n} - \text{F}^{\psi}_{\alpha})^2}{10}} ~;
\end{align}
also see Supplementary information, Table S2 \cite{SIcite}.
Upon comparing the bias in mixed forces ($|\text{F}^{\psi\chi}-\text{F}^{\psi}|$) with the statistical error ($\sigma^{\psi\chi}$), one finds that the largest magnitude of the force bias across temperatures is 1.280 eV/\AA, which is smaller than the largest magnitude of the statistical error, 10.471 eV/\AA. 
Recently, a similar trend between the errors in stochastic forces was seen for the case of an aqueous-solvated peptide system \cite{Shpiro2022}.
 
Figure \ref{fig:fig3} shows a comparison of the components of mixed and stochastic DFT forces (F$^{i, \psi\chi}_{\alpha}$) obtained for several $N_\psi/N_\chi$ with the deterministic Kohn-Sham forces (F$^{i,\psi}_{\alpha}$). The displacement in the mixed forces about the linear curve indicates a statistical error that diminishes with an increasing number of stochastic orbitals at higher temperatures.
At a given temperature, the data points shown in magenta indicate the mixed forces for $N_\psi/N_\chi$ employed in other results presented in this work.
The range of plotted forces is kept constant to compare the spread across temperatures. The forces at $\text{T} = (30, 50)$ eV, are in good agreement between mixed and KS- forces with a standard deviation of 8.647 eV/\AA\, and 10.364 eV/\AA\, respectively. These can be viewed in conjunction with the purely stochastic forces shown in blue.
At lower temperatures, increasing the number of deterministic KS-orbitals reduces the fluctuations in forces, \emph{e.g.}, at $\text{T}=10$ eV, increasing $N_\psi$ from 128/16 (magenta) to 256/16 (yellow) improves the distribution of forces significantly. However, at moderate to high temperatures one would require a drastically large $N_\psi$ to obtain accurate forces, see $\text{T}=50$ eV in Fig. \ref{fig:fig3}. Hence it is advantageous to increase $N_\chi$'s and decrease $N_\psi$ as the temperature increases \cite{WhiteCollins2020}, as evidenced from the forces obtained with $N_\psi/N_\chi =$ 128/16 vs. 16/128 at $\text{T}=50$ eV.

The magnitude of force averaged over all atoms $\langle \text{F}_{\alpha}\rangle$ is computed with mixed and KS-DFT methods, as shown in Fig. \ref{fig:fig4}(a). The error bars denote the standard deviation in the mixed DFT forces, $\sigma^{\psi\chi}$, and the blue-shaded band indicates a thermal fluctuation region described by a Langevin-type fluctuation $\varsigma$, such that
\begin{align}
    m_{\alpha}\ddot{q}_{\alpha} = f_{\alpha} - \gamma_{\alpha}p_{\alpha} + \varsigma\eta_{\alpha}(t) 
\end{align}
where $\varsigma\equiv\sqrt{2m_{\alpha}\gamma_{\alpha}k_BT}$, $(q_{\alpha}, p_{\alpha})$ are the coordinates and momenta of the atoms, $f_{\alpha}$ is the force on the atom, $\gamma_{\alpha}$ is the damping constant, and $\eta_{\alpha}(t)$ is a Gaussian process such that $\left\langle \eta_{\alpha}(t) \right\rangle = 0$ and $\left\langle \eta_{\alpha}(t)\eta_{\alpha'}(t') \right\rangle = \delta_{\alpha\alpha'}\delta(t-t')$.
Langevin molecular dynamics was successfully used in previous studies to investigate forces from stochastic DFT--based simulations \cite{Arnon2020, Shpiro2022}. 
At a given temperature, $\langle \text{F}_{\alpha}\rangle \pm 2\varsigma$ is explicitly computed and then interpolated to yield the thermal-fluctuation band in Fig. \ref{fig:fig4}(a). The averaged mixed DFT forces along with the error bars are contained within the thermal band. This serves to show that the statistical error as captured quantitatively by $\sigma^{\psi\chi}$, and qualitatively in Fig. \ref{fig:fig3}, can be absorbed by the thermal fluctuations in dynamical simulations.
While the statistical fluctuations are contained within $2\varsigma$, the bias in the forces lies within $1\varsigma$ indicating that the accuracy converges faster than precision \cite{Cytter2018}.

In order to investigate the effect of these relatively small biases on observable quantities, we apply mDFT to compute transport properties via molecular dynamics (MD) simulations. We employ an isokinetic \cite{Tuckerman2003}, rather than Langevin, ensemble at each temperature. This is the typical ensemble of choice for WDM transport calculations \cite{White_2017, Ticknor_2016}. The time-dependent free energy and total pressure along with their time-averages and standard deviations are given in Supplementary Information, Table S3, Figs. S5, S6 \cite{SIcite}.
We compute the velocity autocorrelation function (VACF) and self-diffusion coefficient (D) \cite{Ticknor_2016, White_2017, Meyer2014} of carbon at ($\rho$, T) = (3.52 g/cc, 5 eV) with deterministic KS-, mixed and stochastic DFT, see Fig. \ref{fig:fig4}(b). For MD simulations, finite simulation time leads to its own statistical error, in addition to the statistical error due to stochastic calculations in sDFT and mDFT. Estimation of this statistical error depends on the approach to VACF averaging. We see the average mDFT diffusion coefficients falls between 1 and 2 times the statistical error estimate, which is within the range of reasonable estimates, see Supplementary Information for details. The pure sDFT diffusion coefficient falls slightly outside this range, but is still within 10\% of the deterministic case.  
Figure \ref{fig:fig4}(c) shows a comparison of temperature-scaled VACF and D for several T, with the $N_\psi/N_\chi$ for mDFT specified in the key.
The relationship between D and T over a temperature range at any given density was previously investigated for high-Z materials that exhibit multiple ionization states 
\cite{Clerouin2013}.
It was argued that, over a large temperature and density range, the mutually compensating effects of increased ionization and thermal energy result in a constant coupling parameter $\Gamma$, giving rise to a so-called $\Gamma-$plateau which, in turn, affects quantities such as self-diffusion and viscosity. We see that for 1 to 5 eV the change in temperature dominates correlation, leading to an increase in D/T, while for greater than 5 eV the ionization effects become significant leading to a decrease in D/T.  

We have presented the first implementation of the mDFT and sDFT methods within the plane-wave PAW formalism for DFT. The PAW formalism provides a significant acceleration of stochastic DFT methods due to both smaller grids and decreased eigenspectrum range. Additionally it opens the door to efficient, all-electron accuracy, calculations of matter in extreme conditions, as is possible in ambient conditions \cite{Lejaeghere_2016}. We have demonstrated the efficacy of this approach in the simulation of transport properties in isochorically heated warm dense carbon up to 50 eV, observing the crossover from kinetically to Coulomb-dominated correlation effects. Future work will include additional transport studies, and application of the PAW method to time-dependent mDFT and optical response via the Kubo-Greenwood approach. 



\begin{acknowledgments}
This work was supported by the U.S. Department of Energy through the Los Alamos National Laboratory (LANL). Research presented in this article was supported by the Laboratory Directed Research and Development program of LANL, under project number 20210233ER, and Science Campaign 4. We acknowledge the support of the Center for Nonlinear Studies (CNLS). This research used computing resources provided by the LANL Institutional Computing and Advanced Scientific Computing programs. Los Alamos National Laboratory is operated by Triad National Security, LLC, for the National Nuclear Security Administration of U.S. Department of Energy (Contract No. 89233218CNA000001).
\end{acknowledgments}

\bibliography{main}

\begin{thebibliography}{55}%
\makeatletter
\providecommand \@ifxundefined [1]{%
 \@ifx{#1\undefined}
}%
\providecommand \@ifnum [1]{%
 \ifnum #1\expandafter \@firstoftwo
 \else \expandafter \@secondoftwo
 \fi
}%
\providecommand \@ifx [1]{%
 \ifx #1\expandafter \@firstoftwo
 \else \expandafter \@secondoftwo
 \fi
}%
\providecommand \natexlab [1]{#1}%
\providecommand \enquote  [1]{``#1''}%
\providecommand \bibnamefont  [1]{#1}%
\providecommand \bibfnamefont [1]{#1}%
\providecommand \citenamefont [1]{#1}%
\providecommand \href@noop [0]{\@secondoftwo}%
\providecommand \href [0]{\begingroup \@sanitize@url \@href}%
\providecommand \@href[1]{\@@startlink{#1}\@@href}%
\providecommand \@@href[1]{\endgroup#1\@@endlink}%
\providecommand \@sanitize@url [0]{\catcode `\\12\catcode `\$12\catcode
  `\&12\catcode `\#12\catcode `\^12\catcode `\_12\catcode `\%12\relax}%
\providecommand \@@startlink[1]{}%
\providecommand \@@endlink[0]{}%
\providecommand \url  [0]{\begingroup\@sanitize@url \@url }%
\providecommand \@url [1]{\endgroup\@href {#1}{\urlprefix }}%
\providecommand \urlprefix  [0]{URL }%
\providecommand \Eprint [0]{\href }%
\providecommand \doibase [0]{https://doi.org/}%
\providecommand \selectlanguage [0]{\@gobble}%
\providecommand \bibinfo  [0]{\@secondoftwo}%
\providecommand \bibfield  [0]{\@secondoftwo}%
\providecommand \translation [1]{[#1]}%
\providecommand \BibitemOpen [0]{}%
\providecommand \bibitemStop [0]{}%
\providecommand \bibitemNoStop [0]{.\EOS\space}%
\providecommand \EOS [0]{\spacefactor3000\relax}%
\providecommand \BibitemShut  [1]{\csname bibitem#1\endcsname}%
\let\auto@bib@innerbib\@empty
\bibitem [{\citenamefont {{National Academies of Sciences, Engineering, and
  Medicine}}(2018)}]{nas-ess-2018}%
  \BibitemOpen
  \bibfield  {author} {\bibinfo {author} {\bibnamefont {{National Academies of
  Sciences, Engineering, and Medicine}}},\ }\href
  {https://doi.org/{10.17226/25187}} {\emph {\bibinfo {title} {{Exoplanet
  Science Strategy}}}}\ (\bibinfo  {publisher} {The National Academies Press},\
  \bibinfo {address} {Washington, DC},\ \bibinfo {year} {2018})\BibitemShut
  {NoStop}%
\bibitem [{\citenamefont {Bethkenhagen}\ \emph {et~al.}(2017)\citenamefont
  {Bethkenhagen}, \citenamefont {Meyer}, \citenamefont {Hamel}, \citenamefont
  {Nettelmann}, \citenamefont {French}, \citenamefont {Scheibe}, \citenamefont
  {Ticknor}, \citenamefont {Collins}, \citenamefont {Kress}, \citenamefont
  {Fortney},\ and\ \citenamefont {Redmer}}]{bethkenhagen-848-2017}%
  \BibitemOpen
  \bibfield  {author} {\bibinfo {author} {\bibfnamefont {M.}~\bibnamefont
  {Bethkenhagen}}, \bibinfo {author} {\bibfnamefont {E.~R.}\ \bibnamefont
  {Meyer}}, \bibinfo {author} {\bibfnamefont {S.}~\bibnamefont {Hamel}},
  \bibinfo {author} {\bibfnamefont {N.}~\bibnamefont {Nettelmann}}, \bibinfo
  {author} {\bibfnamefont {M.}~\bibnamefont {French}}, \bibinfo {author}
  {\bibfnamefont {L.}~\bibnamefont {Scheibe}}, \bibinfo {author} {\bibfnamefont
  {C.}~\bibnamefont {Ticknor}}, \bibinfo {author} {\bibfnamefont {L.~A.}\
  \bibnamefont {Collins}}, \bibinfo {author} {\bibfnamefont {J.~D.}\
  \bibnamefont {Kress}}, \bibinfo {author} {\bibfnamefont {J.~J.}\ \bibnamefont
  {Fortney}},\ and\ \bibinfo {author} {\bibfnamefont {R.}~\bibnamefont
  {Redmer}},\ }\bibfield  {title} {\bibinfo {title} {{Planetary Ices and the
  Linear Mixing Approximation}},\ }\href
  {https://iopscience.iop.org/article/10.3847/1538-4357/aa8b14} {\bibfield
  {journal} {\bibinfo  {journal} {The Astrophysical Journal}\ }\textbf
  {\bibinfo {volume} {848}},\ \bibinfo {pages} {67} (\bibinfo {year}
  {2017})}\BibitemShut {NoStop}%
\bibitem [{\citenamefont {Teanby}\ \emph {et~al.}(2020)\citenamefont {Teanby},
  \citenamefont {Irwin}, \citenamefont {Moses},\ and\ \citenamefont
  {Helled}}]{Teanby2020}%
  \BibitemOpen
  \bibfield  {author} {\bibinfo {author} {\bibfnamefont {N.~A.}\ \bibnamefont
  {Teanby}}, \bibinfo {author} {\bibfnamefont {P.~G.~J.}\ \bibnamefont
  {Irwin}}, \bibinfo {author} {\bibfnamefont {J.~I.}\ \bibnamefont {Moses}},\
  and\ \bibinfo {author} {\bibfnamefont {R.}~\bibnamefont {Helled}},\
  }\bibfield  {title} {\bibinfo {title} {{Neptune and Uranus: ice or rock
  giants?}},\ }\href {https://doi.org/10.1098/rsta.2019.0489} {\bibfield
  {journal} {\bibinfo  {journal} {Philosophical Transactions of the Royal
  Society A: Mathematical, Physical and Engineering Sciences}\ }\textbf
  {\bibinfo {volume} {378}},\ \bibinfo {pages} {20190489} (\bibinfo {year}
  {2020})}\BibitemShut {NoStop}%
\bibitem [{\citenamefont {Becker}\ \emph {et~al.}(2018)\citenamefont {Becker},
  \citenamefont {Bethkenhagen}, \citenamefont {Kellermann}, \citenamefont
  {Wicht},\ and\ \citenamefont {Redmer}}]{becker-156-2018}%
  \BibitemOpen
  \bibfield  {author} {\bibinfo {author} {\bibfnamefont {A.}~\bibnamefont
  {Becker}}, \bibinfo {author} {\bibfnamefont {M.}~\bibnamefont
  {Bethkenhagen}}, \bibinfo {author} {\bibfnamefont {C.}~\bibnamefont
  {Kellermann}}, \bibinfo {author} {\bibfnamefont {J.}~\bibnamefont {Wicht}},\
  and\ \bibinfo {author} {\bibfnamefont {R.}~\bibnamefont {Redmer}},\
  }\bibfield  {title} {\bibinfo {title} {{Material Properties for the Interiors
  of Massive Giant Planets and Brown Dwarfs}},\ }\href
  {https://iopscience.iop.org/article/10.3847/1538-3881/aad735} {\bibfield
  {journal} {\bibinfo  {journal} {The Astronomical Journal}\ }\textbf {\bibinfo
  {volume} {156}},\ \bibinfo {pages} {149} (\bibinfo {year}
  {2018})}\BibitemShut {NoStop}%
\bibitem [{\citenamefont {Hu}\ \emph {et~al.}(2018)\citenamefont {Hu},
  \citenamefont {Collins}, \citenamefont {Boehly}, \citenamefont {Ding},
  \citenamefont {Radha}, \citenamefont {Goncharov}, \citenamefont {Karasiev},
  \citenamefont {Collins}, \citenamefont {Regan},\ and\ \citenamefont
  {Campbell}}]{hu-25-2018}%
  \BibitemOpen
  \bibfield  {author} {\bibinfo {author} {\bibfnamefont {S.~X.}\ \bibnamefont
  {Hu}}, \bibinfo {author} {\bibfnamefont {L.~A.}\ \bibnamefont {Collins}},
  \bibinfo {author} {\bibfnamefont {T.~R.}\ \bibnamefont {Boehly}}, \bibinfo
  {author} {\bibfnamefont {Y.~H.}\ \bibnamefont {Ding}}, \bibinfo {author}
  {\bibfnamefont {P.~B.}\ \bibnamefont {Radha}}, \bibinfo {author}
  {\bibfnamefont {V.~N.}\ \bibnamefont {Goncharov}}, \bibinfo {author}
  {\bibfnamefont {V.~V.}\ \bibnamefont {Karasiev}}, \bibinfo {author}
  {\bibfnamefont {G.~W.}\ \bibnamefont {Collins}}, \bibinfo {author}
  {\bibfnamefont {S.~P.}\ \bibnamefont {Regan}},\ and\ \bibinfo {author}
  {\bibfnamefont {E.~M.}\ \bibnamefont {Campbell}},\ }\bibfield  {title}
  {\bibinfo {title} {{A review on ab initio studies of static, transport, and
  optical properties of polystyrene under extreme conditions for inertial
  confinement fusion applications}},\ }\href
  {https://aip.scitation.org/doi/full/10.1063/1.5017970} {\bibfield  {journal}
  {\bibinfo  {journal} {Physics of Plasmas}\ }\textbf {\bibinfo {volume}
  {25}},\ \bibinfo {pages} {056306} (\bibinfo {year} {2018})}\BibitemShut
  {NoStop}%
\bibitem [{\citenamefont {Fernandez-Pa\~nella}\ \emph
  {et~al.}(2019)\citenamefont {Fernandez-Pa\~nella}, \citenamefont {Millot},
  \citenamefont {Fratanduono}, \citenamefont {Desjarlais}, \citenamefont
  {Hamel}, \citenamefont {Marshall}, \citenamefont {Erskine}, \citenamefont
  {Sterne}, \citenamefont {Haan}, \citenamefont {Boehly}, \citenamefont
  {Collins}, \citenamefont {Eggert},\ and\ \citenamefont
  {Celliers}}]{fenandez-122-2019}%
  \BibitemOpen
  \bibfield  {author} {\bibinfo {author} {\bibfnamefont {A.}~\bibnamefont
  {Fernandez-Pa\~nella}}, \bibinfo {author} {\bibfnamefont {M.}~\bibnamefont
  {Millot}}, \bibinfo {author} {\bibfnamefont {D.~E.}\ \bibnamefont
  {Fratanduono}}, \bibinfo {author} {\bibfnamefont {M.~P.}\ \bibnamefont
  {Desjarlais}}, \bibinfo {author} {\bibfnamefont {S.}~\bibnamefont {Hamel}},
  \bibinfo {author} {\bibfnamefont {M.~C.}\ \bibnamefont {Marshall}}, \bibinfo
  {author} {\bibfnamefont {D.~J.}\ \bibnamefont {Erskine}}, \bibinfo {author}
  {\bibfnamefont {P.~A.}\ \bibnamefont {Sterne}}, \bibinfo {author}
  {\bibfnamefont {S.}~\bibnamefont {Haan}}, \bibinfo {author} {\bibfnamefont
  {T.~R.}\ \bibnamefont {Boehly}}, \bibinfo {author} {\bibfnamefont {G.~W.}\
  \bibnamefont {Collins}}, \bibinfo {author} {\bibfnamefont {J.~H.}\
  \bibnamefont {Eggert}},\ and\ \bibinfo {author} {\bibfnamefont {P.~M.}\
  \bibnamefont {Celliers}},\ }\bibfield  {title} {\bibinfo {title} {{Shock
  Compression of Liquid Deuterium up to 1 TPa}},\ }\href
  {https://doi.org/10.1103/PhysRevLett.122.255702} {\bibfield  {journal}
  {\bibinfo  {journal} {Phys. Rev. Lett.}\ }\textbf {\bibinfo {volume} {122}},\
  \bibinfo {pages} {255702} (\bibinfo {year} {2019})}\BibitemShut {NoStop}%
\bibitem [{\citenamefont {Bachmann}\ \emph {et~al.}(2022)\citenamefont
  {Bachmann}, \citenamefont {MacLaren}, \citenamefont {Bhandarkar},
  \citenamefont {Briggs}, \citenamefont {Casey}, \citenamefont {Divol},
  \citenamefont {D\"oppner}, \citenamefont {Fittinghoff}, \citenamefont
  {Freeman}, \citenamefont {Haan}, \citenamefont {Hall} \emph
  {et~al.}}]{Bachmann2022}%
  \BibitemOpen
  \bibfield  {author} {\bibinfo {author} {\bibfnamefont {B.}~\bibnamefont
  {Bachmann}}, \bibinfo {author} {\bibfnamefont {S.~A.}\ \bibnamefont
  {MacLaren}}, \bibinfo {author} {\bibfnamefont {S.}~\bibnamefont
  {Bhandarkar}}, \bibinfo {author} {\bibfnamefont {T.}~\bibnamefont {Briggs}},
  \bibinfo {author} {\bibfnamefont {D.}~\bibnamefont {Casey}}, \bibinfo
  {author} {\bibfnamefont {L.}~\bibnamefont {Divol}}, \bibinfo {author}
  {\bibfnamefont {T.}~\bibnamefont {D\"oppner}}, \bibinfo {author}
  {\bibfnamefont {D.}~\bibnamefont {Fittinghoff}}, \bibinfo {author}
  {\bibfnamefont {M.}~\bibnamefont {Freeman}}, \bibinfo {author} {\bibfnamefont
  {S.}~\bibnamefont {Haan}}, \bibinfo {author} {\bibfnamefont {G.~N.}\
  \bibnamefont {Hall}}, \emph {et~al.},\ }\bibfield  {title} {\bibinfo {title}
  {{Measurement of Dark Ice-Ablator Mix in Inertial Confinement Fusion}},\
  }\href {https://doi.org/10.1103/PhysRevLett.129.275001} {\bibfield  {journal}
  {\bibinfo  {journal} {Phys. Rev. Lett.}\ }\textbf {\bibinfo {volume} {129}},\
  \bibinfo {pages} {275001} (\bibinfo {year} {2022})}\BibitemShut {NoStop}%
\bibitem [{\citenamefont {Millot}\ \emph {et~al.}(2018)\citenamefont {Millot},
  \citenamefont {Hamel}, \citenamefont {Rygg}, \citenamefont {Celliers},
  \citenamefont {Collins}, \citenamefont {Coppari}, \citenamefont
  {Fratanduono}, \citenamefont {Jeanloz}, \citenamefont {Swift},\ and\
  \citenamefont {Eggert}}]{millot-14-2018}%
  \BibitemOpen
  \bibfield  {author} {\bibinfo {author} {\bibfnamefont {M.}~\bibnamefont
  {Millot}}, \bibinfo {author} {\bibfnamefont {S.}~\bibnamefont {Hamel}},
  \bibinfo {author} {\bibfnamefont {J.~R.}\ \bibnamefont {Rygg}}, \bibinfo
  {author} {\bibfnamefont {P.~M.}\ \bibnamefont {Celliers}}, \bibinfo {author}
  {\bibfnamefont {G.~W.}\ \bibnamefont {Collins}}, \bibinfo {author}
  {\bibfnamefont {F.}~\bibnamefont {Coppari}}, \bibinfo {author} {\bibfnamefont
  {D.~E.}\ \bibnamefont {Fratanduono}}, \bibinfo {author} {\bibfnamefont
  {R.}~\bibnamefont {Jeanloz}}, \bibinfo {author} {\bibfnamefont {D.~C.}\
  \bibnamefont {Swift}},\ and\ \bibinfo {author} {\bibfnamefont {J.~H.}\
  \bibnamefont {Eggert}},\ }\bibfield  {title} {\bibinfo {title} {{Experimental
  evidence for superionic water ice using shock compression}},\ }\href
  {https://www.nature.com/articles/s41567-017-0017-4} {\bibfield  {journal}
  {\bibinfo  {journal} {Nature Physics}\ }\textbf {\bibinfo {volume} {14}},\
  \bibinfo {pages} {297} (\bibinfo {year} {2018})}\BibitemShut {NoStop}%
\bibitem [{\citenamefont {Cheng}\ \emph {et~al.}(2021)\citenamefont {Cheng},
  \citenamefont {Bethkenhagen}, \citenamefont {Pickard},\ and\ \citenamefont
  {Hamel}}]{cheng-17-2021}%
  \BibitemOpen
  \bibfield  {author} {\bibinfo {author} {\bibfnamefont {B.}~\bibnamefont
  {Cheng}}, \bibinfo {author} {\bibfnamefont {M.}~\bibnamefont {Bethkenhagen}},
  \bibinfo {author} {\bibfnamefont {C.~J.}\ \bibnamefont {Pickard}},\ and\
  \bibinfo {author} {\bibfnamefont {S.}~\bibnamefont {Hamel}},\ }\bibfield
  {title} {\bibinfo {title} {{Phase behaviours of superionic water at planetary
  conditions}},\ }\href {https://www.nature.com/articles/s41567-021-01334-9}
  {\bibfield  {journal} {\bibinfo  {journal} {Nature Physics}\ }\textbf
  {\bibinfo {volume} {17}},\ \bibinfo {pages} {1228} (\bibinfo {year}
  {2021})}\BibitemShut {NoStop}%
\bibitem [{\citenamefont {Gao}\ \emph {et~al.}(2022)\citenamefont {Gao},
  \citenamefont {Liu}, \citenamefont {Shi}, \citenamefont {Pan}, \citenamefont
  {Huang}, \citenamefont {Lu}, \citenamefont {Wang}, \citenamefont {Xing},\
  and\ \citenamefont {Sun}}]{gao-128-2022}%
  \BibitemOpen
  \bibfield  {author} {\bibinfo {author} {\bibfnamefont {H.}~\bibnamefont
  {Gao}}, \bibinfo {author} {\bibfnamefont {C.}~\bibnamefont {Liu}}, \bibinfo
  {author} {\bibfnamefont {J.}~\bibnamefont {Shi}}, \bibinfo {author}
  {\bibfnamefont {S.}~\bibnamefont {Pan}}, \bibinfo {author} {\bibfnamefont
  {T.}~\bibnamefont {Huang}}, \bibinfo {author} {\bibfnamefont
  {X.}~\bibnamefont {Lu}}, \bibinfo {author} {\bibfnamefont {H.-T.}\
  \bibnamefont {Wang}}, \bibinfo {author} {\bibfnamefont {D.}~\bibnamefont
  {Xing}},\ and\ \bibinfo {author} {\bibfnamefont {J.}~\bibnamefont {Sun}},\
  }\bibfield  {title} {\bibinfo {title} {{Superionic Silica-Water and
  Silica-Hydrogen Compounds in the Deep Interiors of Uranus and Neptune}},\
  }\href {https://doi.org/10.1103/PhysRevLett.128.035702} {\bibfield  {journal}
  {\bibinfo  {journal} {Phys. Rev. Lett.}\ }\textbf {\bibinfo {volume} {128}},\
  \bibinfo {pages} {035702} (\bibinfo {year} {2022})}\BibitemShut {NoStop}%
\bibitem [{\citenamefont {Cheng}\ \emph {et~al.}(2022)\citenamefont {Cheng},
  \citenamefont {Hamel},\ and\ \citenamefont
  {Bethkenhagen}}]{cheng-arXiv-2023}%
  \BibitemOpen
  \bibfield  {author} {\bibinfo {author} {\bibfnamefont {B.}~\bibnamefont
  {Cheng}}, \bibinfo {author} {\bibfnamefont {S.}~\bibnamefont {Hamel}},\ and\
  \bibinfo {author} {\bibfnamefont {M.}~\bibnamefont {Bethkenhagen}},\ }\href
  {https://doi.org/10.48550/arxiv.2207.02927} {\bibinfo {title} {{Diamond
  formation from hydrocarbon mixtures in planets}}} (\bibinfo {year}
  {2022})\BibitemShut {NoStop}%
\bibitem [{\citenamefont {Ross}(1981)}]{ross-292-1981}%
  \BibitemOpen
  \bibfield  {author} {\bibinfo {author} {\bibfnamefont {M.}~\bibnamefont
  {Ross}},\ }\bibfield  {title} {\bibinfo {title} {{The ice layer in Uranus and
  Neptune -- diamonds in the sky?}},\ }\href
  {https://www.nature.com/articles/292435a0} {\bibfield  {journal} {\bibinfo
  {journal} {Nature}\ }\textbf {\bibinfo {volume} {292}},\ \bibinfo {pages}
  {435} (\bibinfo {year} {1981})}\BibitemShut {NoStop}%
\bibitem [{\citenamefont {White}\ \emph {et~al.}(2018)\citenamefont {White},
  \citenamefont {Certik}, \citenamefont {Ding}, \citenamefont {Hu},\ and\
  \citenamefont {Collins}}]{white-98-2018}%
  \BibitemOpen
  \bibfield  {author} {\bibinfo {author} {\bibfnamefont {A.~J.}\ \bibnamefont
  {White}}, \bibinfo {author} {\bibfnamefont {O.}~\bibnamefont {Certik}},
  \bibinfo {author} {\bibfnamefont {Y.~H.}\ \bibnamefont {Ding}}, \bibinfo
  {author} {\bibfnamefont {S.~X.}\ \bibnamefont {Hu}},\ and\ \bibinfo {author}
  {\bibfnamefont {L.~A.}\ \bibnamefont {Collins}},\ }\bibfield  {title}
  {\bibinfo {title} {{Time-dependent orbital-free density functional theory for
  electronic stopping power: Comparison to the Mermin-Kohn-Sham theory at high
  temperatures}},\ }\href {https://doi.org/10.1103/PhysRevB.98.144302}
  {\bibfield  {journal} {\bibinfo  {journal} {Phys. Rev. B}\ }\textbf {\bibinfo
  {volume} {98}},\ \bibinfo {pages} {144302} (\bibinfo {year}
  {2018})}\BibitemShut {NoStop}%
\bibitem [{\citenamefont {Ding}\ \emph {et~al.}(2018)\citenamefont {Ding},
  \citenamefont {White}, \citenamefont {Hu}, \citenamefont {Certik},\ and\
  \citenamefont {Collins}}]{ding-121-2018}%
  \BibitemOpen
  \bibfield  {author} {\bibinfo {author} {\bibfnamefont {Y.~H.}\ \bibnamefont
  {Ding}}, \bibinfo {author} {\bibfnamefont {A.~J.}\ \bibnamefont {White}},
  \bibinfo {author} {\bibfnamefont {S.~X.}\ \bibnamefont {Hu}}, \bibinfo
  {author} {\bibfnamefont {O.}~\bibnamefont {Certik}},\ and\ \bibinfo {author}
  {\bibfnamefont {L.~A.}\ \bibnamefont {Collins}},\ }\bibfield  {title}
  {\bibinfo {title} {{Ab Initio Studies on the Stopping Power of Warm Dense
  Matter with Time-Dependent Orbital-Free Density Functional Theory}},\ }\href
  {https://doi.org/10.1103/PhysRevLett.121.145001} {\bibfield  {journal}
  {\bibinfo  {journal} {Phys. Rev. Lett.}\ }\textbf {\bibinfo {volume} {121}},\
  \bibinfo {pages} {145001} (\bibinfo {year} {2018})}\BibitemShut {NoStop}%
\bibitem [{\citenamefont {Ahrer}\ \emph {et~al.}(2022)\citenamefont {Ahrer},
  \citenamefont {Alderson}, \citenamefont {Batalha}, \citenamefont {Batalha},
  \citenamefont {Bean}, \citenamefont {Beatty}, \citenamefont {Bell},
  \citenamefont {Benneke}, \citenamefont {Berta-Thompson}, \citenamefont
  {Carter}, \citenamefont {Crossfield} \emph {et~al.}}]{ahrer-arX-2022}%
  \BibitemOpen
  \bibfield  {author} {\bibinfo {author} {\bibfnamefont {E.-M.}\ \bibnamefont
  {Ahrer}}, \bibinfo {author} {\bibfnamefont {L.}~\bibnamefont {Alderson}},
  \bibinfo {author} {\bibfnamefont {N.~M.}\ \bibnamefont {Batalha}}, \bibinfo
  {author} {\bibfnamefont {N.~E.}\ \bibnamefont {Batalha}}, \bibinfo {author}
  {\bibfnamefont {J.~L.}\ \bibnamefont {Bean}}, \bibinfo {author}
  {\bibfnamefont {T.~G.}\ \bibnamefont {Beatty}}, \bibinfo {author}
  {\bibfnamefont {T.~J.}\ \bibnamefont {Bell}}, \bibinfo {author}
  {\bibfnamefont {B.}~\bibnamefont {Benneke}}, \bibinfo {author} {\bibfnamefont
  {Z.~K.}\ \bibnamefont {Berta-Thompson}}, \bibinfo {author} {\bibfnamefont
  {A.~L.}\ \bibnamefont {Carter}}, \bibinfo {author} {\bibfnamefont {I.~J.~M.}\
  \bibnamefont {Crossfield}}, \emph {et~al.},\ }\bibfield  {title} {\bibinfo
  {title} {Identification of carbon dioxide in an exoplanet atmosphere},\
  }\href {https://www.nature.com/articles/s41586-022-05269-w} {\bibfield
  {journal} {\bibinfo  {journal} {Nature}\ } (\bibinfo {year}
  {2022})}\BibitemShut {NoStop}%
\bibitem [{\citenamefont {Dietrich}\ \emph {et~al.}(2022)\citenamefont
  {Dietrich}, \citenamefont {Kumar}, \citenamefont {Poser}, \citenamefont
  {French}, \citenamefont {Nettelmann}, \citenamefont {Redmer},\ and\
  \citenamefont {Wicht}}]{detrich-517-2022}%
  \BibitemOpen
  \bibfield  {author} {\bibinfo {author} {\bibfnamefont {W.}~\bibnamefont
  {Dietrich}}, \bibinfo {author} {\bibfnamefont {S.}~\bibnamefont {Kumar}},
  \bibinfo {author} {\bibfnamefont {A.~J.}\ \bibnamefont {Poser}}, \bibinfo
  {author} {\bibfnamefont {M.}~\bibnamefont {French}}, \bibinfo {author}
  {\bibfnamefont {N.}~\bibnamefont {Nettelmann}}, \bibinfo {author}
  {\bibfnamefont {R.}~\bibnamefont {Redmer}},\ and\ \bibinfo {author}
  {\bibfnamefont {J.}~\bibnamefont {Wicht}},\ }\bibfield  {title} {\bibinfo
  {title} {{Magnetic induction processes in hot Jupiters, application to
  KELT-9b}},\ }\href {https://doi.org/10.1093/mnras/stac2849} {\bibfield
  {journal} {\bibinfo  {journal} {Monthly Notices of the Royal Astronomical
  Society}\ }\textbf {\bibinfo {volume} {517}},\ \bibinfo {pages} {3113}
  (\bibinfo {year} {2022})}\BibitemShut {NoStop}%
\bibitem [{\citenamefont {Miyazaki}\ and\ \citenamefont
  {Stevenson}(2022)}]{miyazaki-3-2022}%
  \BibitemOpen
  \bibfield  {author} {\bibinfo {author} {\bibfnamefont {Y.}~\bibnamefont
  {Miyazaki}}\ and\ \bibinfo {author} {\bibfnamefont {D.~J.}\ \bibnamefont
  {Stevenson}},\ }\bibfield  {title} {\bibinfo {title} {{A Subsurface Magma
  Ocean on Io: Exploring the Steady State of Partially Molten Planetary
  Bodies}},\ }\href {https://iopscience.iop.org/article/10.3847/PSJ/ac9cd1}
  {\bibfield  {journal} {\bibinfo  {journal} {The Planetary Science Journal}\
  }\textbf {\bibinfo {volume} {3}},\ \bibinfo {pages} {256} (\bibinfo {year}
  {2022})}\BibitemShut {NoStop}%
\bibitem [{\citenamefont {Liu}\ \emph {et~al.}(2019)\citenamefont {Liu},
  \citenamefont {Hori}, \citenamefont {M{\"u}ller}, \citenamefont {Zheng},
  \citenamefont {Helled}, \citenamefont {Lin},\ and\ \citenamefont
  {Isella}}]{liu-542-2019}%
  \BibitemOpen
  \bibfield  {author} {\bibinfo {author} {\bibfnamefont {S.-F.}\ \bibnamefont
  {Liu}}, \bibinfo {author} {\bibfnamefont {Y.}~\bibnamefont {Hori}}, \bibinfo
  {author} {\bibfnamefont {S.}~\bibnamefont {M{\"u}ller}}, \bibinfo {author}
  {\bibfnamefont {X.}~\bibnamefont {Zheng}}, \bibinfo {author} {\bibfnamefont
  {R.}~\bibnamefont {Helled}}, \bibinfo {author} {\bibfnamefont
  {D.}~\bibnamefont {Lin}},\ and\ \bibinfo {author} {\bibfnamefont
  {A.}~\bibnamefont {Isella}},\ }\bibfield  {title} {\bibinfo {title} {{The
  formation of Jupiter's diluted core by a giant impact}},\ }\href
  {https://www.nature.com/articles/s41586-019-1470-2} {\bibfield  {journal}
  {\bibinfo  {journal} {Nature}\ }\textbf {\bibinfo {volume} {572}},\ \bibinfo
  {pages} {355} (\bibinfo {year} {2019})}\BibitemShut {NoStop}%
\bibitem [{\citenamefont {Zylstra}\ \emph {et~al.}(2022)\citenamefont
  {Zylstra}, \citenamefont {Hurricane}, \citenamefont {Callahan}, \citenamefont
  {Kritcher}, \citenamefont {Ralph}, \citenamefont {Robey}, \citenamefont
  {Ross}, \citenamefont {Young}, \citenamefont {Baker}, \citenamefont {Casey},
  \citenamefont {D{\"o}ppner} \emph {et~al.}}]{zylstra-601-2022}%
  \BibitemOpen
  \bibfield  {author} {\bibinfo {author} {\bibfnamefont {A.~B.}\ \bibnamefont
  {Zylstra}}, \bibinfo {author} {\bibfnamefont {O.~A.}\ \bibnamefont
  {Hurricane}}, \bibinfo {author} {\bibfnamefont {D.~A.}\ \bibnamefont
  {Callahan}}, \bibinfo {author} {\bibfnamefont {A.~L.}\ \bibnamefont
  {Kritcher}}, \bibinfo {author} {\bibfnamefont {J.~E.}\ \bibnamefont {Ralph}},
  \bibinfo {author} {\bibfnamefont {H.~F.}\ \bibnamefont {Robey}}, \bibinfo
  {author} {\bibfnamefont {J.~S.}\ \bibnamefont {Ross}}, \bibinfo {author}
  {\bibfnamefont {C.~V.}\ \bibnamefont {Young}}, \bibinfo {author}
  {\bibfnamefont {K.~L.}\ \bibnamefont {Baker}}, \bibinfo {author}
  {\bibfnamefont {D.~T.}\ \bibnamefont {Casey}}, \bibinfo {author}
  {\bibfnamefont {T.}~\bibnamefont {D{\"o}ppner}}, \emph {et~al.},\ }\bibfield
  {title} {\bibinfo {title} {{Burning plasma achieved in inertial fusion}},\
  }\href {https://www.nature.com/articles/s41586-021-04281-w} {\bibfield
  {journal} {\bibinfo  {journal} {Nature}\ }\textbf {\bibinfo {volume} {601}},\
  \bibinfo {pages} {542} (\bibinfo {year} {2022})}\BibitemShut {NoStop}%
\bibitem [{\citenamefont {Kohn}\ and\ \citenamefont
  {Sham}(1965)}]{kohn-140-1965}%
  \BibitemOpen
  \bibfield  {author} {\bibinfo {author} {\bibfnamefont {W.}~\bibnamefont
  {Kohn}}\ and\ \bibinfo {author} {\bibfnamefont {L.~J.}\ \bibnamefont
  {Sham}},\ }\bibfield  {title} {\bibinfo {title} {{Self-Consistent Equations
  Including Exchange and Correlation Effects}},\ }\href
  {https://doi.org/10.1103/PhysRev.140.A1133} {\bibfield  {journal} {\bibinfo
  {journal} {Phys. Rev.}\ }\textbf {\bibinfo {volume} {140}},\ \bibinfo {pages}
  {A1133} (\bibinfo {year} {1965})}\BibitemShut {NoStop}%
\bibitem [{\citenamefont {Pribram-Jones}\ \emph {et~al.}(2014)\citenamefont
  {Pribram-Jones}, \citenamefont {Pittalis}, \citenamefont {Gross},\ and\
  \citenamefont {Burke}}]{burke2014}%
  \BibitemOpen
  \bibfield  {author} {\bibinfo {author} {\bibfnamefont {A.}~\bibnamefont
  {Pribram-Jones}}, \bibinfo {author} {\bibfnamefont {S.}~\bibnamefont
  {Pittalis}}, \bibinfo {author} {\bibfnamefont {E.~K.~U.}\ \bibnamefont
  {Gross}},\ and\ \bibinfo {author} {\bibfnamefont {K.}~\bibnamefont {Burke}},\
  }\bibfield  {title} {\bibinfo {title} {{Thermal Density Functional Theory in
  Context}},\ }in\ \href@noop {} {\emph {\bibinfo {booktitle} {Frontiers and
  Challenges in Warm Dense Matter}}},\ \bibinfo {editor} {edited by\ \bibinfo
  {editor} {\bibfnamefont {F.}~\bibnamefont {Graziani}}, \bibinfo {editor}
  {\bibfnamefont {M.~P.}\ \bibnamefont {Desjarlais}}, \bibinfo {editor}
  {\bibfnamefont {R.}~\bibnamefont {Redmer}},\ and\ \bibinfo {editor}
  {\bibfnamefont {S.~B.}\ \bibnamefont {Trickey}}}\ (\bibinfo  {publisher}
  {Springer International Publishing},\ \bibinfo {address} {Cham},\ \bibinfo
  {year} {2014})\ pp.\ \bibinfo {pages} {25--60}\BibitemShut {NoStop}%
\bibitem [{\citenamefont {Bonitz}\ \emph {et~al.}(2020)\citenamefont {Bonitz},
  \citenamefont {Dornheim}, \citenamefont {Moldabekov}, \citenamefont {Zhang},
  \citenamefont {Hamann}, \citenamefont {K\"{a}hlert}, \citenamefont {Filinov},
  \citenamefont {Ramakrishna},\ and\ \citenamefont {Vorberger}}]{Bonitz2020}%
  \BibitemOpen
  \bibfield  {author} {\bibinfo {author} {\bibfnamefont {M.}~\bibnamefont
  {Bonitz}}, \bibinfo {author} {\bibfnamefont {T.}~\bibnamefont {Dornheim}},
  \bibinfo {author} {\bibfnamefont {Z.~A.}\ \bibnamefont {Moldabekov}},
  \bibinfo {author} {\bibfnamefont {S.}~\bibnamefont {Zhang}}, \bibinfo
  {author} {\bibfnamefont {P.}~\bibnamefont {Hamann}}, \bibinfo {author}
  {\bibfnamefont {H.}~\bibnamefont {K\"{a}hlert}}, \bibinfo {author}
  {\bibfnamefont {A.}~\bibnamefont {Filinov}}, \bibinfo {author} {\bibfnamefont
  {K.}~\bibnamefont {Ramakrishna}},\ and\ \bibinfo {author} {\bibfnamefont
  {J.}~\bibnamefont {Vorberger}},\ }\bibfield  {title} {\bibinfo {title} {{Ab
  initio simulation of warm dense matter}},\ }\href
  {https://doi.org/10.1063/1.5143225} {\bibfield  {journal} {\bibinfo
  {journal} {Physics of Plasmas}\ }\textbf {\bibinfo {volume} {27}},\ \bibinfo
  {pages} {042710} (\bibinfo {year} {2020})}\BibitemShut {NoStop}%
\bibitem [{\citenamefont {Ceperley}(1995)}]{ceperley-67-1995}%
  \BibitemOpen
  \bibfield  {author} {\bibinfo {author} {\bibfnamefont {D.~M.}\ \bibnamefont
  {Ceperley}},\ }\bibfield  {title} {\bibinfo {title} {{Path integrals in the
  theory of condensed helium}},\ }\href
  {https://doi.org/10.1103/RevModPhys.67.279} {\bibfield  {journal} {\bibinfo
  {journal} {Rev. Mod. Phys.}\ }\textbf {\bibinfo {volume} {67}},\ \bibinfo
  {pages} {279} (\bibinfo {year} {1995})}\BibitemShut {NoStop}%
\bibitem [{\citenamefont {Driver}\ and\ \citenamefont
  {Militzer}(2012)}]{militzer-108-2012}%
  \BibitemOpen
  \bibfield  {author} {\bibinfo {author} {\bibfnamefont {K.~P.}\ \bibnamefont
  {Driver}}\ and\ \bibinfo {author} {\bibfnamefont {B.}~\bibnamefont
  {Militzer}},\ }\bibfield  {title} {\bibinfo {title} {{All-Electron Path
  Integral Monte Carlo Simulations of Warm Dense Matter: Application to Water
  and Carbon Plasmas}},\ }\href
  {https://doi.org/10.1103/PhysRevLett.108.115502} {\bibfield  {journal}
  {\bibinfo  {journal} {Phys. Rev. Lett.}\ }\textbf {\bibinfo {volume} {108}},\
  \bibinfo {pages} {115502} (\bibinfo {year} {2012})}\BibitemShut {NoStop}%
\bibitem [{\citenamefont {Deringer}\ \emph {et~al.}(2019)\citenamefont
  {Deringer}, \citenamefont {Caro},\ and\ \citenamefont
  {Csányi}}]{deringer-31-2019}%
  \BibitemOpen
  \bibfield  {author} {\bibinfo {author} {\bibfnamefont {V.~L.}\ \bibnamefont
  {Deringer}}, \bibinfo {author} {\bibfnamefont {M.~A.}\ \bibnamefont {Caro}},\
  and\ \bibinfo {author} {\bibfnamefont {G.}~\bibnamefont {Csányi}},\
  }\bibfield  {title} {\bibinfo {title} {{Machine Learning Interatomic
  Potentials as Emerging Tools for Materials Science}},\ }\href
  {https://doi.org/https://doi.org/10.1002/adma.201902765} {\bibfield
  {journal} {\bibinfo  {journal} {Advanced Materials}\ }\textbf {\bibinfo
  {volume} {31}},\ \bibinfo {pages} {1902765} (\bibinfo {year}
  {2019})}\BibitemShut {NoStop}%
\bibitem [{\citenamefont {Li}\ \emph {et~al.}(2022)\citenamefont {Li},
  \citenamefont {Oganov}, \citenamefont {Cui}, \citenamefont {Zhou},
  \citenamefont {Dong},\ and\ \citenamefont {Wang}}]{li-128-2022}%
  \BibitemOpen
  \bibfield  {author} {\bibinfo {author} {\bibfnamefont {H.-F.}\ \bibnamefont
  {Li}}, \bibinfo {author} {\bibfnamefont {A.~R.}\ \bibnamefont {Oganov}},
  \bibinfo {author} {\bibfnamefont {H.}~\bibnamefont {Cui}}, \bibinfo {author}
  {\bibfnamefont {X.-F.}\ \bibnamefont {Zhou}}, \bibinfo {author}
  {\bibfnamefont {X.}~\bibnamefont {Dong}},\ and\ \bibinfo {author}
  {\bibfnamefont {H.-T.}\ \bibnamefont {Wang}},\ }\bibfield  {title} {\bibinfo
  {title} {{Ultrahigh-Pressure Magnesium Hydrosilicates as Reservoirs of Water
  in Early Earth}},\ }\href {https://doi.org/10.1103/PhysRevLett.128.035703}
  {\bibfield  {journal} {\bibinfo  {journal} {Phys. Rev. Lett.}\ }\textbf
  {\bibinfo {volume} {128}},\ \bibinfo {pages} {035703} (\bibinfo {year}
  {2022})}\BibitemShut {NoStop}%
\bibitem [{\citenamefont {Blanchet}\ \emph {et~al.}(2020)\citenamefont
  {Blanchet}, \citenamefont {Torrent},\ and\ \citenamefont
  {Cl{\'e}rouin}}]{Blanchet_2020}%
  \BibitemOpen
  \bibfield  {author} {\bibinfo {author} {\bibfnamefont {A.}~\bibnamefont
  {Blanchet}}, \bibinfo {author} {\bibfnamefont {M.}~\bibnamefont {Torrent}},\
  and\ \bibinfo {author} {\bibfnamefont {J.}~\bibnamefont {Cl{\'e}rouin}},\
  }\bibfield  {title} {\bibinfo {title} {{Requirements for very high
  temperature Kohn--Sham DFT simulations and how to bypass them}},\ }\href
  {https://aip.scitation.org/doi/10.1063/5.0016538} {\bibfield  {journal}
  {\bibinfo  {journal} {Physics of Plasmas}\ }\textbf {\bibinfo {volume}
  {27}},\ \bibinfo {pages} {122706} (\bibinfo {year} {2020})}\BibitemShut
  {NoStop}%
\bibitem [{\citenamefont {Lambert}\ \emph {et~al.}(2006)\citenamefont
  {Lambert}, \citenamefont {Cl{\'e}rouin},\ and\ \citenamefont
  {Mazevet}}]{Lambert_2006}%
  \BibitemOpen
  \bibfield  {author} {\bibinfo {author} {\bibfnamefont {F.}~\bibnamefont
  {Lambert}}, \bibinfo {author} {\bibfnamefont {J.}~\bibnamefont
  {Cl{\'e}rouin}},\ and\ \bibinfo {author} {\bibfnamefont {S.}~\bibnamefont
  {Mazevet}},\ }\bibfield  {title} {\bibinfo {title} {{Structural and dynamical
  properties of hot dense matter by a Thomas-Fermi-Dirac molecular dynamics}},\
  }\href {https://iopscience.iop.org/article/10.1209/epl/i2006-10184-7}
  {\bibfield  {journal} {\bibinfo  {journal} {Europhysics Letters}\ }\textbf
  {\bibinfo {volume} {75}},\ \bibinfo {pages} {681} (\bibinfo {year}
  {2006})}\BibitemShut {NoStop}%
\bibitem [{\citenamefont {Ticknor}\ \emph {et~al.}(2016)\citenamefont
  {Ticknor}, \citenamefont {Kress}, \citenamefont {Collins}, \citenamefont
  {Cl\'erouin}, \citenamefont {Arnault},\ and\ \citenamefont
  {Decoster}}]{Ticknor_2016}%
  \BibitemOpen
  \bibfield  {author} {\bibinfo {author} {\bibfnamefont {C.}~\bibnamefont
  {Ticknor}}, \bibinfo {author} {\bibfnamefont {J.~D.}\ \bibnamefont {Kress}},
  \bibinfo {author} {\bibfnamefont {L.~A.}\ \bibnamefont {Collins}}, \bibinfo
  {author} {\bibfnamefont {J.}~\bibnamefont {Cl\'erouin}}, \bibinfo {author}
  {\bibfnamefont {P.}~\bibnamefont {Arnault}},\ and\ \bibinfo {author}
  {\bibfnamefont {A.}~\bibnamefont {Decoster}},\ }\bibfield  {title} {\bibinfo
  {title} {{Transport properties of an asymmetric mixture in the dense plasma
  regime}},\ }\href {https://doi.org/10.1103/PhysRevE.93.063208} {\bibfield
  {journal} {\bibinfo  {journal} {Phys. Rev. E}\ }\textbf {\bibinfo {volume}
  {93}},\ \bibinfo {pages} {063208} (\bibinfo {year} {2016})}\BibitemShut
  {NoStop}%
\bibitem [{\citenamefont {White}\ \emph {et~al.}(2017)\citenamefont {White},
  \citenamefont {Collins}, \citenamefont {Kress}, \citenamefont {Ticknor},
  \citenamefont {Cl\'erouin}, \citenamefont {Arnault},\ and\ \citenamefont
  {Desbiens}}]{White_2017}%
  \BibitemOpen
  \bibfield  {author} {\bibinfo {author} {\bibfnamefont {A.~J.}\ \bibnamefont
  {White}}, \bibinfo {author} {\bibfnamefont {L.~A.}\ \bibnamefont {Collins}},
  \bibinfo {author} {\bibfnamefont {J.~D.}\ \bibnamefont {Kress}}, \bibinfo
  {author} {\bibfnamefont {C.}~\bibnamefont {Ticknor}}, \bibinfo {author}
  {\bibfnamefont {J.}~\bibnamefont {Cl\'erouin}}, \bibinfo {author}
  {\bibfnamefont {P.}~\bibnamefont {Arnault}},\ and\ \bibinfo {author}
  {\bibfnamefont {N.}~\bibnamefont {Desbiens}},\ }\bibfield  {title} {\bibinfo
  {title} {{Correlation and transport properties for mixtures at constant
  pressure and temperature}},\ }\href
  {https://doi.org/10.1103/PhysRevE.95.063202} {\bibfield  {journal} {\bibinfo
  {journal} {Phys. Rev. E}\ }\textbf {\bibinfo {volume} {95}},\ \bibinfo
  {pages} {063202} (\bibinfo {year} {2017})}\BibitemShut {NoStop}%
\bibitem [{\citenamefont {Fabian}\ \emph {et~al.}(2019)\citenamefont {Fabian},
  \citenamefont {Shpiro}, \citenamefont {Rabani}, \citenamefont {Neuhauser},\
  and\ \citenamefont {Baer}}]{Fabian2019}%
  \BibitemOpen
  \bibfield  {author} {\bibinfo {author} {\bibfnamefont {M.~D.}\ \bibnamefont
  {Fabian}}, \bibinfo {author} {\bibfnamefont {B.}~\bibnamefont {Shpiro}},
  \bibinfo {author} {\bibfnamefont {E.}~\bibnamefont {Rabani}}, \bibinfo
  {author} {\bibfnamefont {D.}~\bibnamefont {Neuhauser}},\ and\ \bibinfo
  {author} {\bibfnamefont {R.}~\bibnamefont {Baer}},\ }\bibfield  {title}
  {\bibinfo {title} {{Stochastic density functional theory}},\ }\href
  {https://doi.org/https://doi.org/10.1002/wcms.1412} {\bibfield  {journal}
  {\bibinfo  {journal} {WIREs Computational Molecular Science}\ }\textbf
  {\bibinfo {volume} {9}},\ \bibinfo {pages} {e1412} (\bibinfo {year}
  {2019})}\BibitemShut {NoStop}%
\bibitem [{\citenamefont {Cytter}\ \emph {et~al.}(2018)\citenamefont {Cytter},
  \citenamefont {Rabani}, \citenamefont {Neuhauser},\ and\ \citenamefont
  {Baer}}]{Cytter2018}%
  \BibitemOpen
  \bibfield  {author} {\bibinfo {author} {\bibfnamefont {Y.}~\bibnamefont
  {Cytter}}, \bibinfo {author} {\bibfnamefont {E.}~\bibnamefont {Rabani}},
  \bibinfo {author} {\bibfnamefont {D.}~\bibnamefont {Neuhauser}},\ and\
  \bibinfo {author} {\bibfnamefont {R.}~\bibnamefont {Baer}},\ }\bibfield
  {title} {\bibinfo {title} {{Stochastic density functional theory at finite
  temperatures}},\ }\href {https://doi.org/10.1103/PhysRevB.97.115207}
  {\bibfield  {journal} {\bibinfo  {journal} {Phys. Rev. B}\ }\textbf {\bibinfo
  {volume} {97}},\ \bibinfo {pages} {115207} (\bibinfo {year}
  {2018})}\BibitemShut {NoStop}%
\bibitem [{\citenamefont {White}\ and\ \citenamefont
  {Collins}(2020)}]{WhiteCollins2020}%
  \BibitemOpen
  \bibfield  {author} {\bibinfo {author} {\bibfnamefont {A.~J.}\ \bibnamefont
  {White}}\ and\ \bibinfo {author} {\bibfnamefont {L.~A.}\ \bibnamefont
  {Collins}},\ }\bibfield  {title} {\bibinfo {title} {{Fast and Universal
  Kohn-Sham Density Functional Theory Algorithm for Warm Dense Matter to Hot
  Dense Plasma}},\ }\href {https://doi.org/10.1103/PhysRevLett.125.055002}
  {\bibfield  {journal} {\bibinfo  {journal} {Phys. Rev. Lett.}\ }\textbf
  {\bibinfo {volume} {125}},\ \bibinfo {pages} {055002} (\bibinfo {year}
  {2020})}\BibitemShut {NoStop}%
\bibitem [{\citenamefont {Laasonen}\ \emph {et~al.}(1993)\citenamefont
  {Laasonen}, \citenamefont {Pasquarello}, \citenamefont {Car}, \citenamefont
  {Lee},\ and\ \citenamefont {Vanderbilt}}]{Laasonen_1993}%
  \BibitemOpen
  \bibfield  {author} {\bibinfo {author} {\bibfnamefont {K.}~\bibnamefont
  {Laasonen}}, \bibinfo {author} {\bibfnamefont {A.}~\bibnamefont
  {Pasquarello}}, \bibinfo {author} {\bibfnamefont {R.}~\bibnamefont {Car}},
  \bibinfo {author} {\bibfnamefont {C.}~\bibnamefont {Lee}},\ and\ \bibinfo
  {author} {\bibfnamefont {D.}~\bibnamefont {Vanderbilt}},\ }\bibfield  {title}
  {\bibinfo {title} {{Car-Parrinello molecular dynamics with Vanderbilt
  ultrasoft pseudopotentials}},\ }\href
  {https://doi.org/10.1103/PhysRevB.47.10142} {\bibfield  {journal} {\bibinfo
  {journal} {Phys. Rev. B}\ }\textbf {\bibinfo {volume} {47}},\ \bibinfo
  {pages} {10142} (\bibinfo {year} {1993})}\BibitemShut {NoStop}%
\bibitem [{\citenamefont {Bl\"ochl}(1994)}]{Blochl_1994}%
  \BibitemOpen
  \bibfield  {author} {\bibinfo {author} {\bibfnamefont {P.~E.}\ \bibnamefont
  {Bl\"ochl}},\ }\bibfield  {title} {\bibinfo {title} {{Projector
  augmented-wave method}},\ }\href {https://doi.org/10.1103/PhysRevB.50.17953}
  {\bibfield  {journal} {\bibinfo  {journal} {Phys. Rev. B}\ }\textbf {\bibinfo
  {volume} {50}},\ \bibinfo {pages} {17953} (\bibinfo {year}
  {1994})}\BibitemShut {NoStop}%
\bibitem [{\citenamefont {Kresse}\ and\ \citenamefont
  {Joubert}(1999)}]{Kresse_1999}%
  \BibitemOpen
  \bibfield  {author} {\bibinfo {author} {\bibfnamefont {G.}~\bibnamefont
  {Kresse}}\ and\ \bibinfo {author} {\bibfnamefont {D.}~\bibnamefont
  {Joubert}},\ }\bibfield  {title} {\bibinfo {title} {{From ultrasoft
  pseudopotentials to the projector augmented-wave method}},\ }\href
  {https://doi.org/10.1103/PhysRevB.59.1758} {\bibfield  {journal} {\bibinfo
  {journal} {Phys. Rev. B}\ }\textbf {\bibinfo {volume} {59}},\ \bibinfo
  {pages} {1758} (\bibinfo {year} {1999})}\BibitemShut {NoStop}%
\bibitem [{\citenamefont {Lejaeghere}\ \emph {et~al.}(2016)\citenamefont
  {Lejaeghere}, \citenamefont {Bihlmayer}, \citenamefont {Bj\"{o}rkman},
  \citenamefont {Blaha}, \citenamefont {Bl\"{u}gel}, \citenamefont {Blum},
  \citenamefont {Caliste}, \citenamefont {Castelli}, \citenamefont {Clark},
  \citenamefont {Corso}, \citenamefont {de~Gironcoli} \emph
  {et~al.}}]{Lejaeghere_2016}%
  \BibitemOpen
  \bibfield  {author} {\bibinfo {author} {\bibfnamefont {K.}~\bibnamefont
  {Lejaeghere}}, \bibinfo {author} {\bibfnamefont {G.}~\bibnamefont
  {Bihlmayer}}, \bibinfo {author} {\bibfnamefont {T.}~\bibnamefont
  {Bj\"{o}rkman}}, \bibinfo {author} {\bibfnamefont {P.}~\bibnamefont {Blaha}},
  \bibinfo {author} {\bibfnamefont {S.}~\bibnamefont {Bl\"{u}gel}}, \bibinfo
  {author} {\bibfnamefont {V.}~\bibnamefont {Blum}}, \bibinfo {author}
  {\bibfnamefont {D.}~\bibnamefont {Caliste}}, \bibinfo {author} {\bibfnamefont
  {I.~E.}\ \bibnamefont {Castelli}}, \bibinfo {author} {\bibfnamefont {S.~J.}\
  \bibnamefont {Clark}}, \bibinfo {author} {\bibfnamefont {A.~D.}\ \bibnamefont
  {Corso}}, \bibinfo {author} {\bibfnamefont {S.}~\bibnamefont {de~Gironcoli}},
  \emph {et~al.},\ }\bibfield  {title} {\bibinfo {title} {{Reproducibility in
  density functional theory calculations of solids}},\ }\href
  {https://doi.org/10.1126/science.aad3000} {\bibfield  {journal} {\bibinfo
  {journal} {Science}\ }\textbf {\bibinfo {volume} {351}},\ \bibinfo {pages}
  {aad3000} (\bibinfo {year} {2016})}\BibitemShut {NoStop}%
\bibitem [{\citenamefont {Mermin}(1965)}]{Mermin1965}%
  \BibitemOpen
  \bibfield  {author} {\bibinfo {author} {\bibfnamefont {N.~D.}\ \bibnamefont
  {Mermin}},\ }\bibfield  {title} {\bibinfo {title} {{Thermal Properties of the
  Inhomogeneous Electron Gas}},\ }\href
  {https://doi.org/10.1103/PhysRev.137.A1441} {\bibfield  {journal} {\bibinfo
  {journal} {Phys. Rev.}\ }\textbf {\bibinfo {volume} {137}},\ \bibinfo {pages}
  {A1441} (\bibinfo {year} {1965})}\BibitemShut {NoStop}%
\bibitem [{\citenamefont {Baer}\ \emph {et~al.}(2013)\citenamefont {Baer},
  \citenamefont {Neuhauser},\ and\ \citenamefont {Rabani}}]{Baer2013}%
  \BibitemOpen
  \bibfield  {author} {\bibinfo {author} {\bibfnamefont {R.}~\bibnamefont
  {Baer}}, \bibinfo {author} {\bibfnamefont {D.}~\bibnamefont {Neuhauser}},\
  and\ \bibinfo {author} {\bibfnamefont {E.}~\bibnamefont {Rabani}},\
  }\bibfield  {title} {\bibinfo {title} {{Self-Averaging Stochastic Kohn-Sham
  Density-Functional Theory}},\ }\href
  {https://doi.org/10.1103/PhysRevLett.111.106402} {\bibfield  {journal}
  {\bibinfo  {journal} {Phys. Rev. Lett.}\ }\textbf {\bibinfo {volume} {111}},\
  \bibinfo {pages} {106402} (\bibinfo {year} {2013})}\BibitemShut {NoStop}%
\bibitem [{\citenamefont {Neuhauser}\ \emph {et~al.}(2014)\citenamefont
  {Neuhauser}, \citenamefont {Baer},\ and\ \citenamefont
  {Rabani}}]{Neuhauser2014}%
  \BibitemOpen
  \bibfield  {author} {\bibinfo {author} {\bibfnamefont {D.}~\bibnamefont
  {Neuhauser}}, \bibinfo {author} {\bibfnamefont {R.}~\bibnamefont {Baer}},\
  and\ \bibinfo {author} {\bibfnamefont {E.}~\bibnamefont {Rabani}},\
  }\bibfield  {title} {\bibinfo {title} {{Communication: Embedded fragment
  stochastic density functional theory}},\ }\href
  {https://doi.org/10.1063/1.4890651} {\bibfield  {journal} {\bibinfo
  {journal} {The Journal of Chemical Physics}\ }\textbf {\bibinfo {volume}
  {141}},\ \bibinfo {pages} {041102} (\bibinfo {year} {2014})}\BibitemShut
  {NoStop}%
\bibitem [{\citenamefont {Baer}\ \emph {et~al.}(2022)\citenamefont {Baer},
  \citenamefont {Neuhauser},\ and\ \citenamefont {Rabani}}]{Baer2022}%
  \BibitemOpen
  \bibfield  {author} {\bibinfo {author} {\bibfnamefont {R.}~\bibnamefont
  {Baer}}, \bibinfo {author} {\bibfnamefont {D.}~\bibnamefont {Neuhauser}},\
  and\ \bibinfo {author} {\bibfnamefont {E.}~\bibnamefont {Rabani}},\
  }\bibfield  {title} {\bibinfo {title} {{Stochastic Vector Techniques in
  Ground-State Electronic Structure}},\ }\href
  {https://doi.org/10.1146/annurev-physchem-090519-045916} {\bibfield
  {journal} {\bibinfo  {journal} {Annual Review of Physical Chemistry}\
  }\textbf {\bibinfo {volume} {73}},\ \bibinfo {pages} {255} (\bibinfo {year}
  {2022})}\BibitemShut {NoStop}%
\bibitem [{\citenamefont {Hutchinson}(1990)}]{Hutchinson1990}%
  \BibitemOpen
  \bibfield  {author} {\bibinfo {author} {\bibfnamefont {M.}~\bibnamefont
  {Hutchinson}},\ }\bibfield  {title} {\bibinfo {title} {{A stochastic
  estimator of the trace of the influence matrix for laplacian smoothing
  splines}},\ }\href {https://doi.org/10.1080/03610919008812866} {\bibfield
  {journal} {\bibinfo  {journal} {Communications in Statistics - Simulation and
  Computation}\ }\textbf {\bibinfo {volume} {19}},\ \bibinfo {pages} {433}
  (\bibinfo {year} {1990})}\BibitemShut {NoStop}%
\bibitem [{SIc()}]{SIcite}%
  \BibitemOpen
  \href@noop {} {\ }\bibinfo {note} {{\!\!Supplementary Information for
  ``Stochastic and Mixed Kohn Sham Density Functional Theory within the
  projector augmented wave formalism for simulation of warm dense
  matter."}}\BibitemShut {NoStop}%
\bibitem [{\citenamefont {Li}\ and\ \citenamefont
  {Neuhauser}(2020)}]{Neuhauser2020}%
  \BibitemOpen
  \bibfield  {author} {\bibinfo {author} {\bibfnamefont {W.}~\bibnamefont
  {Li}}\ and\ \bibinfo {author} {\bibfnamefont {D.}~\bibnamefont {Neuhauser}},\
  }\bibfield  {title} {\bibinfo {title} {{Real-space orthogonal
  projector-augmented-wave method}},\ }\href
  {https://doi.org/10.1103/PhysRevB.102.195118} {\bibfield  {journal} {\bibinfo
   {journal} {Phys. Rev. B}\ }\textbf {\bibinfo {volume} {102}},\ \bibinfo
  {pages} {195118} (\bibinfo {year} {2020})}\BibitemShut {NoStop}%
\bibitem [{\citenamefont {Levitt}\ and\ \citenamefont
  {Torrent}(2015)}]{LEVITT201598}%
  \BibitemOpen
  \bibfield  {author} {\bibinfo {author} {\bibfnamefont {A.}~\bibnamefont
  {Levitt}}\ and\ \bibinfo {author} {\bibfnamefont {M.}~\bibnamefont
  {Torrent}},\ }\bibfield  {title} {\bibinfo {title} {{Parallel eigensolvers in
  plane-wave Density Functional Theory}},\ }\href
  {https://www.sciencedirect.com/science/article/pii/S0010465514003531}
  {\bibfield  {journal} {\bibinfo  {journal} {Computer Physics Communications}\
  }\textbf {\bibinfo {volume} {187}},\ \bibinfo {pages} {98} (\bibinfo {year}
  {2015})}\BibitemShut {NoStop}%
\bibitem [{\citenamefont {Torrent}\ \emph {et~al.}(2008)\citenamefont
  {Torrent}, \citenamefont {Jollet}, \citenamefont {Bottin}, \citenamefont
  {Z{\'e}rah},\ and\ \citenamefont {Gonze}}]{Torrent_2008}%
  \BibitemOpen
  \bibfield  {author} {\bibinfo {author} {\bibfnamefont {M.}~\bibnamefont
  {Torrent}}, \bibinfo {author} {\bibfnamefont {F.}~\bibnamefont {Jollet}},
  \bibinfo {author} {\bibfnamefont {F.}~\bibnamefont {Bottin}}, \bibinfo
  {author} {\bibfnamefont {G.}~\bibnamefont {Z{\'e}rah}},\ and\ \bibinfo
  {author} {\bibfnamefont {X.}~\bibnamefont {Gonze}},\ }\bibfield  {title}
  {\bibinfo {title} {{Implementation of the projector augmented-wave method in
  the ABINIT code: Application to the study of iron under pressure}},\
  }\href@noop {} {\bibfield  {journal} {\bibinfo  {journal} {Computational
  Materials Science}\ }\textbf {\bibinfo {volume} {42}},\ \bibinfo {pages}
  {337} (\bibinfo {year} {2008})}\BibitemShut {NoStop}%
\bibitem [{\citenamefont {Bennett}\ \emph {et~al.}(1978)\citenamefont
  {Bennett}, \citenamefont {Johnson}, \citenamefont {Kerley},\ and\
  \citenamefont {Rood}}]{sesame}%
  \BibitemOpen
  \bibfield  {author} {\bibinfo {author} {\bibfnamefont {B.}~\bibnamefont
  {Bennett}}, \bibinfo {author} {\bibfnamefont {J.}~\bibnamefont {Johnson}},
  \bibinfo {author} {\bibfnamefont {G.}~\bibnamefont {Kerley}},\ and\ \bibinfo
  {author} {\bibfnamefont {G.}~\bibnamefont {Rood}},\ }\bibfield  {title}
  {\bibinfo {title} {{Recent Developments in the Sesame Equation-Of-State
  Library}},\ }\bibfield  {journal} {\bibinfo  {journal} {Los Alamos National
  Laboratory LA-7130}\ }\href {https://doi.org/10.2172/5150206}
  {10.2172/5150206} (\bibinfo {year} {1978})\BibitemShut {NoStop}%
\bibitem [{\citenamefont {Rangel}\ \emph {et~al.}(2016)\citenamefont {Rangel},
  \citenamefont {Caliste}, \citenamefont {Genovese},\ and\ \citenamefont
  {Torrent}}]{libpaw2016}%
  \BibitemOpen
  \bibfield  {author} {\bibinfo {author} {\bibfnamefont {T.}~\bibnamefont
  {Rangel}}, \bibinfo {author} {\bibfnamefont {D.}~\bibnamefont {Caliste}},
  \bibinfo {author} {\bibfnamefont {L.}~\bibnamefont {Genovese}},\ and\
  \bibinfo {author} {\bibfnamefont {M.}~\bibnamefont {Torrent}},\ }\bibfield
  {title} {\bibinfo {title} {{A wavelet-based Projector Augmented-Wave (PAW)
  method: Reaching frozen-core all-electron precision with a systematic,
  adaptive and localized wavelet basis set}},\ }\href
  {https://doi.org/https://doi.org/10.1016/j.cpc.2016.06.012} {\bibfield
  {journal} {\bibinfo  {journal} {Computer Physics Communications}\ }\textbf
  {\bibinfo {volume} {208}},\ \bibinfo {pages} {1} (\bibinfo {year}
  {2016})}\BibitemShut {NoStop}%
\bibitem [{\citenamefont {Romero}\ \emph {et~al.}(2020)\citenamefont {Romero},
  \citenamefont {Allan}, \citenamefont {Amadon}, \citenamefont {Antonius},
  \citenamefont {Applencourt}, \citenamefont {Baguet}, \citenamefont {Bieder},
  \citenamefont {Bottin}, \citenamefont {Bouchet}, \citenamefont {Bousquet},
  \citenamefont {Bruneval} \emph {et~al.}}]{abinit2020}%
  \BibitemOpen
  \bibfield  {author} {\bibinfo {author} {\bibfnamefont {A.~H.}\ \bibnamefont
  {Romero}}, \bibinfo {author} {\bibfnamefont {D.~C.}\ \bibnamefont {Allan}},
  \bibinfo {author} {\bibfnamefont {B.}~\bibnamefont {Amadon}}, \bibinfo
  {author} {\bibfnamefont {G.}~\bibnamefont {Antonius}}, \bibinfo {author}
  {\bibfnamefont {T.}~\bibnamefont {Applencourt}}, \bibinfo {author}
  {\bibfnamefont {L.}~\bibnamefont {Baguet}}, \bibinfo {author} {\bibfnamefont
  {J.}~\bibnamefont {Bieder}}, \bibinfo {author} {\bibfnamefont
  {F.}~\bibnamefont {Bottin}}, \bibinfo {author} {\bibfnamefont
  {J.}~\bibnamefont {Bouchet}}, \bibinfo {author} {\bibfnamefont
  {E.}~\bibnamefont {Bousquet}}, \bibinfo {author} {\bibfnamefont
  {F.}~\bibnamefont {Bruneval}}, \emph {et~al.},\ }\bibfield  {title} {\bibinfo
  {title} {{ABINIT: Overview and focus on selected capabilities}},\ }\href
  {https://doi.org/10.1063/1.5144261} {\bibfield  {journal} {\bibinfo
  {journal} {The Journal of Chemical Physics}\ }\textbf {\bibinfo {volume}
  {152}},\ \bibinfo {pages} {124102} (\bibinfo {year} {2020})}\BibitemShut
  {NoStop}%
\bibitem [{\citenamefont {Lehtola}\ \emph {et~al.}(2018)\citenamefont
  {Lehtola}, \citenamefont {Steigemann}, \citenamefont {Oliveira},\ and\
  \citenamefont {Marques}}]{libxc2018}%
  \BibitemOpen
  \bibfield  {author} {\bibinfo {author} {\bibfnamefont {S.}~\bibnamefont
  {Lehtola}}, \bibinfo {author} {\bibfnamefont {C.}~\bibnamefont {Steigemann}},
  \bibinfo {author} {\bibfnamefont {M.~J.}\ \bibnamefont {Oliveira}},\ and\
  \bibinfo {author} {\bibfnamefont {M.~A.}\ \bibnamefont {Marques}},\
  }\bibfield  {title} {\bibinfo {title} {{Recent developments in LIBXC -- A
  comprehensive library of functionals for density functional theory}},\ }\href
  {https://doi.org/https://doi.org/10.1016/j.softx.2017.11.002} {\bibfield
  {journal} {\bibinfo  {journal} {SoftwareX}\ }\textbf {\bibinfo {volume}
  {7}},\ \bibinfo {pages} {1} (\bibinfo {year} {2018})}\BibitemShut {NoStop}%
\bibitem [{\citenamefont {Arnon}\ \emph {et~al.}(2020)\citenamefont {Arnon},
  \citenamefont {Rabani}, \citenamefont {Neuhauser},\ and\ \citenamefont
  {Baer}}]{Arnon2020}%
  \BibitemOpen
  \bibfield  {author} {\bibinfo {author} {\bibfnamefont {E.}~\bibnamefont
  {Arnon}}, \bibinfo {author} {\bibfnamefont {E.}~\bibnamefont {Rabani}},
  \bibinfo {author} {\bibfnamefont {D.}~\bibnamefont {Neuhauser}},\ and\
  \bibinfo {author} {\bibfnamefont {R.}~\bibnamefont {Baer}},\ }\bibfield
  {title} {\bibinfo {title} {{Efficient Langevin dynamics for ``noisy"
  forces}},\ }\href {https://doi.org/10.1063/5.0004954} {\bibfield  {journal}
  {\bibinfo  {journal} {The Journal of Chemical Physics}\ }\textbf {\bibinfo
  {volume} {152}},\ \bibinfo {pages} {161103} (\bibinfo {year}
  {2020})}\BibitemShut {NoStop}%
\bibitem [{\citenamefont {Shpiro}\ \emph {et~al.}(2022)\citenamefont {Shpiro},
  \citenamefont {Fabian}, \citenamefont {Rabani},\ and\ \citenamefont
  {Baer}}]{Shpiro2022}%
  \BibitemOpen
  \bibfield  {author} {\bibinfo {author} {\bibfnamefont {B.}~\bibnamefont
  {Shpiro}}, \bibinfo {author} {\bibfnamefont {M.~D.}\ \bibnamefont {Fabian}},
  \bibinfo {author} {\bibfnamefont {E.}~\bibnamefont {Rabani}},\ and\ \bibinfo
  {author} {\bibfnamefont {R.}~\bibnamefont {Baer}},\ }\bibfield  {title}
  {\bibinfo {title} {{Forces from Stochastic Density Functional Theory under
  Nonorthogonal Atom-Centered Basis Sets}},\ }\href
  {https://doi.org/10.1021/acs.jctc.1c00794} {\bibfield  {journal} {\bibinfo
  {journal} {Journal of Chemical Theory and Computation}\ }\textbf {\bibinfo
  {volume} {18}},\ \bibinfo {pages} {1458} (\bibinfo {year}
  {2022})}\BibitemShut {NoStop}%
\bibitem [{\citenamefont {Minary}\ \emph {et~al.}(2003)\citenamefont {Minary},
  \citenamefont {Martyna},\ and\ \citenamefont {Tuckerman}}]{Tuckerman2003}%
  \BibitemOpen
  \bibfield  {author} {\bibinfo {author} {\bibfnamefont {P.}~\bibnamefont
  {Minary}}, \bibinfo {author} {\bibfnamefont {G.~J.}\ \bibnamefont
  {Martyna}},\ and\ \bibinfo {author} {\bibfnamefont {M.~E.}\ \bibnamefont
  {Tuckerman}},\ }\bibfield  {title} {\bibinfo {title} {{Algorithms and novel
  applications based on the isokinetic ensemble. II. Ab initio molecular
  dynamics}},\ }\href {https://doi.org/10.1063/1.1534583} {\bibfield  {journal}
  {\bibinfo  {journal} {The Journal of Chemical Physics}\ }\textbf {\bibinfo
  {volume} {118}},\ \bibinfo {pages} {2527} (\bibinfo {year}
  {2003})}\BibitemShut {NoStop}%
\bibitem [{\citenamefont {Meyer}\ \emph {et~al.}(2014)\citenamefont {Meyer},
  \citenamefont {Kress}, \citenamefont {Collins},\ and\ \citenamefont
  {Ticknor}}]{Meyer2014}%
  \BibitemOpen
  \bibfield  {author} {\bibinfo {author} {\bibfnamefont {E.~R.}\ \bibnamefont
  {Meyer}}, \bibinfo {author} {\bibfnamefont {J.~D.}\ \bibnamefont {Kress}},
  \bibinfo {author} {\bibfnamefont {L.~A.}\ \bibnamefont {Collins}},\ and\
  \bibinfo {author} {\bibfnamefont {C.}~\bibnamefont {Ticknor}},\ }\bibfield
  {title} {\bibinfo {title} {{Effect of correlation on viscosity and diffusion
  in molecular-dynamics simulations}},\ }\href
  {https://doi.org/10.1103/PhysRevE.90.043101} {\bibfield  {journal} {\bibinfo
  {journal} {Phys. Rev. E}\ }\textbf {\bibinfo {volume} {90}},\ \bibinfo
  {pages} {043101} (\bibinfo {year} {2014})}\BibitemShut {NoStop}%
\bibitem [{\citenamefont {Cl\'erouin}\ \emph {et~al.}(2013)\citenamefont
  {Cl\'erouin}, \citenamefont {Robert}, \citenamefont {Arnault}, \citenamefont
  {Kress},\ and\ \citenamefont {Collins}}]{Clerouin2013}%
  \BibitemOpen
  \bibfield  {author} {\bibinfo {author} {\bibfnamefont {J.}~\bibnamefont
  {Cl\'erouin}}, \bibinfo {author} {\bibfnamefont {G.}~\bibnamefont {Robert}},
  \bibinfo {author} {\bibfnamefont {P.}~\bibnamefont {Arnault}}, \bibinfo
  {author} {\bibfnamefont {J.~D.}\ \bibnamefont {Kress}},\ and\ \bibinfo
  {author} {\bibfnamefont {L.~A.}\ \bibnamefont {Collins}},\ }\bibfield
  {title} {\bibinfo {title} {{Behavior of the coupling parameter under
  isochoric heating in a high-$Z$ plasma}},\ }\href
  {https://doi.org/10.1103/PhysRevE.87.061101} {\bibfield  {journal} {\bibinfo
  {journal} {Phys. Rev. E}\ }\textbf {\bibinfo {volume} {87}},\ \bibinfo
  {pages} {061101} (\bibinfo {year} {2013})}\BibitemShut {NoStop}%
\end{thebibliography}%


%

\end{document}